# Multiscale modeling of regularly staggered carbon fibers embedded in nano-reinforced composites

## S. I. Kundalwal and S. A. Meguid*

Mechanics and Aerospace Design Laboratory, Department of Mechanical and Industrial Engineering, University of Toronto, Toronto, Ontario M5S 3G8, Canada

## Abstract

This article deals with the multiscale modeling of stress transfer characteristics of nano-reinforced polymer composite reinforced with regularly staggered carbon fibers. The distinctive feature of construction of nano-reinforced composite is such that the microscale carbon fibers are packed in hexagonal array in the carbon nanotube reinforced polymer matrix (CNRP). We considered three different cases of CNRP, in which carbon nanotubes (CNTs) are: (i) aligned along the direction of carbon fiber, (ii) aligned radially to the axis of carbon fiber, and (iii) randomly dispersed. Accordingly, multiscale models were developed. First, molecular dynamics (MD) simulations and then Mori-Tanaka technique were used to estimate the effective elastic properties of CNRP. Second, a micromechanical three-phase shear lag model was developed considering the staggering effect of microscale fibers and the application of radial loads on the cylindrical representative volume element (RVE) of nano-reinforced composite. Our results reveal that the stress transfer characteristics of the nano-reinforced composite are significantly improved by controlling the CNT morphology, particularly, when they are randomly dispersed around the microscale fiber. The results from the developed shear lag model were also validated with the finite element shear lag simulations and found to be in good agreement.

# 1. Introduction

Fiber-matrix interfacial properties significantly affect the structural performance of composites subjected to external loads. The ability to tailor fiber-matrix interfacial properties is essential to ensure efficient load transfer from matrix to the fibers, which help to reduce stress concentrations and damage, and enhance overall mechanical behavior of resulting composite [1]. Several experimental and analytical techniques have been developed thus far to gain insights into the basic mechanisms dominating the fiber-matrix interfacial characteristics because the strength and toughness of the composite is largely dependent on the nature of fiber-matrix interface. To characterize stress transfer mechanisms, the pull-out test or shear lag model is typically employed. A significant number of analytical and computational two- and three-phase shear lag models have been developed to better understand the stress transfer mechanisms across the fiber-matrix interface [2-6]. These models differ in terms of whether the interphase between the fiber and the matrix is considered or not, and whether we are concerned with long fiber or short fiber composites. In the case of three-phase shear lag model, a thin interphase formed as a result of chemical interactions between the fiber and the matrix, is considered. The chemical composition of such interphase differs from both the fiber and matrix constituents but its mechanical properties lie between these two [7-9], and such nanoscale interphase has a marginal influence on the bulk elastic properties of a composite. On the other hand, a third phase (that is, relatively thick interphase) made of different material can be engineered between the fiber and the matrix [6, 10]. Such microscale interphase strongly influences the mechanical and interfacial properties of a composite, where the reinforcing effect is related to the interfacial adhesion strength, and the interphase serving to inhibit crack propagation or as mechanical damping elements [see [11] and the references therein].

CNTs [12] have been emerged as the ideal candidates for multifarious applications due to their remarkable elastic and physical properties. A CNT can be viewed as a hollow seamless cylinder formed by rolling a graphene sheet. A significant number of experimental and numerical studies have been carried out to estimate the elastic properties of CNTs [13-15], and reported that the axial Young's modulus of CNTs is in the TeraPascal range. The quest for utilizing such exceptional elastic properties of CNTs has led to the opening of an emerging area of research concerned with the development of two-phase CNT-reinforced nanocomposites [9, 16-23]. However, the addition of CNTs in polymer matrix does not always result in improved effective properties of the two-phase nanocomposites. Several important factors, such as agglomeration, aggregation and waviness of CNTs, and difficulty in manufacturing also play a significant role [17]. These difficulties can be alleviated by using CNTs as secondary reinforcements in a three-phase CNT-reinforced composite. Extensive research has been dedicated to the introduction of CNTs as the modifiers to the conventional composites in order to enhance their multifunctional properties. For example, [24] reported an approach to the development of advanced structural composites based on engineered CNT-microscale fiber reinforcement; the CNT-carbon fabric-epoxy composites showed ∼30% enhancement of the interlaminar shear strength as compared to that of microscale fiber-epoxy composites. [25] grew aligned CNTs (A-CNTs) on the circumferential surfaces of microfibers to reinforce the matrix and reported the improvement in multifunctional properties. [26] fabricated unidirectional composite in which CNTs were directly grown on the circumferential surfaces of conventional microscale fibers. [27] fabricated the carbon fiber reinforced composite incorporating functionalized CNTs in the epoxy matrix; as a consequence, they observed significant improvements in tensile strength, stiffness and resistance to failure due to cyclic loadings. [11] deposited CNTs on the circumferential surfaces of electrically insulated glass fiber surfaces. According to their fragmentation test results, the incorporation of an interphase with a small number of CNTs around the fiber, remarkably improved the interfacial shear strength of the fiber-epoxy composite. The functionalized CNTs were incorporated by [28] at the fiber/fabric−matrix interfaces of a carbon fiber-epoxy composite; their study showed improvements in the tensile strength and stiffness, and resistance to tension−tension fatigue damage due to the created CNT- reinforced region at the fiber/fabric−matrix interfaces. [29] developed a shear lag model for a ceramic matrix composite containing wavy, finite length nanofibers having a statistical distribution of strengths as a function of all the material parameters including morphology. A numerical method is proposed by [30] to theoretically investigate the pull-out of a hybrid fiber coated with CNTs. They developed two-step finite element (FE) approach: a single CNT pull-out from the matrix at microscale and the pull-out of the hybrid fiber at macroscale. Their numerical results indicate that the apparent interfacial shear strength of the hybrid fiber and the specific pull-out energy are significantly increased due to the additional bonding of the CNT−matrix interface. A beneficial interfacial effect of the presence of CNTs on the circumferential surface of the microscale fiber samples is demonstrated by [31] resulting in an increase in the interlaminar shear strength (>30 MPa) over uncoated samples. This increase is attributed to an enhanced contact between the resin and the fibers due to an increased surface area as a result of the CNTs. Recently, current authors (Kundalwal et al., 2014) investigated the stress transfer characteristics of a novel hybrid hierarchical nanocomposite in which the carbon fibers augmented with CNTs on their circumferential surfaces are interlaced in the polymer matrix. They developed micromechanics models to determine the elastic properties of CNRP phase and the stress transfer characteristics of hybrid composite. Two types of CNT morphologies are studied by [32]: CNTs deposited in fiber coatings and CNTs grown on circumferential surfaces of fibers. In the former case CNTs are tangentially aligned along the fiber surface and are radially aligned in the latter case. Most recently, the improved mechanical properties and stress transfer behavior of multiscale composite containing nano- and micro-scale reinforcements is reported by Kundalwal and Kumar [33]. They developed three-phase pull-out model to analyze the stress transfer characteristics within a single cylindrical RVE, neglecting the staggering effect of adjacent RVEs.

Findings in the literature indicate that the use of CNTs and conventional microscale fibers together, as multiscale reinforcements, significantly improve the overall properties of resulting hybrid composites, which are unachievable in conventional composites. It is well known that damage initiation is progressive with the applied load and that the small crack at the fiber-matrix interface may reduce the fatigue life of composites. By toughening the interfacial fiber-matrix region with nano-fillers, we can enhance the long-term performance as well as damage initiation threshold of conventional composites. Indeed, this concept can be exploited to grade the bulk matrix properties around the conventional fibers, which may eventually improve the stress transfer behaviour of multiscale composite. To the best of our knowledge, there has been no multiscale model to study the stress transfer characteristics of nano-reinforced composite containing orthotropic micro- and nano-scale fillers. Existing shear lag models are insufficient to accurately describe the stress transfer mechanisms because they model the interactions among constituents at different scales through analytical or numerical micromechanical techniques. This provides the motivation behind the current study. Therefore, a more comprehensive multiscale model was advanced in this study to investigate the stress transfer mechanisms of nano-reinforced composite, in which the effects of different scales of constituents were taken into account. First, we developed multiscale model to determine the orthotropic elastic properties of CNRP through MD simulations in conjunction with the Mori-Tanaka model. The resulting intermediate CNRP phase, containing CNTs and epoxy, was considered as an interphase. Then the determined elastic moduli of the CNRP were used in the development of three-phase shear lag model accounting staggering effect of adjacent RVEs. Particular attention was paid to investigate (i) the influence of orientation of CNTs, and (ii) the effect of variation of the axial and lateral spacing between the adjacent microscale fibers on the stress transfer characteristics of the composite.

## 2. Multiscale modeling

For most multiscale composites, mechanical response and fracture behavior arise from the properties of its different constituents at each level and the interaction between these constituents. Thus, different multiscale models have been developed over the last decade to predict the overall properties of composites at the microscale level [34, 35] and the references therein]. Here, multiscale modeling of a nano-reinforced composite was achieved in two consecutive steps: (i) orthotropic elastic properties of the nano-fiber made of a single CNT and the epoxy matrix were determined through MD simulations; (ii) the determined elastic properties of the nano-fiber and epoxy were used to determine the bulk elastic properties of the CNRP using Mori-Tanaka model. Figure 1 shows the steps involved in the hierarchical multiscale model.

### 2.1. Molecular modeling

This section describes the procedure for building a series of MD models for the epoxy and the nano-fiber. The technique for creating an epoxy and nano-fiber is described first, followed by the MD simulations for determining the isotropic elastic properties of the epoxy material and the orthotropic elastic properties of the nano-fiber. All MD simulations runs were conducted with large-scale atomic/molecular massively parallel simulator [LAMMPS; [36]]. The consistent valence force field [CVFF; [37]] was used to describe the atomic interactions between different atoms. A very efficient conjugate gradient algorithm was used to minimize the strain energy as a function of the displacement of the MD systems while the velocity Verlet algorithm was used to integrate the equations of motion in all MD simulations. Periodic boundary conditions were applied to the MD unit cell faces. Determination of the elastic properties of the pure epoxy and the nano-fiber was accomplished by straining the MD unit cells using constant-strain energy minimization. The averaged stresses on the MD unit cell was defined in the form of virial stress [38]; as follows

$$\sigma = \frac{1}{V} \sum_{i=1}^{N} \left( \frac{m_i}{2} v_i^2 + F_i r_i \right) \tag{1}$$

where V is the volume of the unit cell; $v_i$, $m_i$, $r_i$ and $F_i$ are the velocity, mass, position and force acting on the ith atom, respectively.

### 2.1.1. Modeling of EPON 862-DETDA epoxy

Thermosetting polymers are the matrices of choice for structural composites due to their high stiffness, strength, and creep and thermal resistance over the thermoplastic polymers [39]. Many thermosetting polymers are formed by mixing a resin and a curing agent. We used epoxy material based on EPON 862 resin and Diethylene Toluene Diamine (DETDA) curing agent to form a crosslinked structure, which is typically used in the aerospace industry. The molecular structures of these two monomers are shown in Fig. 2(a). To achieve the crosslinking structure, the potential reactive sites in the epoxy resin can be activated by hydrating the epoxy oxygen atoms at the ends of the molecule, see Fig. 2(b). The EPON 862-DETDA weight ratio was set to 2:1 to obtain the best elastic properties [40]. The initial MD unit cell, consisting of both activated epoxy (100 molecules of EPON 862) and a curing agent (50 molecules of DETDA), was built using the PACKMOL software [41], as shown in Fig. 3. The polymerization process usually occurs into two main stages: pre-curing equilibration and curing of the polymer network. These main steps involved in determining the elastic moduli of pure epoxy are described as follows:

**Step 1 (pre-curing equilibration):** The initial MD unit cell was compressed gradually through several steps from its initial size of 50 Å × 50 Å × 50 Å to the targeted dimensions of 39 Å × 39 Å × 39 Å. The details of the MD unit cell are summarized in Table 1. At each stage, the atoms coordinates were remapped to fit inside the compressed box then a minimization simulation was carried out to relax their coordinates. The system was considered to be optimized once the change in the total potential energy of the system between subsequent steps was less than $1.0 \times 10^{-10}$ kcal/mol [34]. The optimized system was then equilibrated at room temperature in the NVT ensemble over 100 ps using a time step of 1 fs.

**Step 2 (curing):** After the MD unit cell was fully equilibrated in step 1, the polymerization and crosslinking were simulated by allowing chemical reactions between reactive atoms. Chemical reactions were simulated in a stepwise manner using a criterion based on atomic distances and the type of chemical primary or secondary amine reactions as described elsewhere [40]. The distance between all pairs of reactive C–N atoms were computed and new bonds were created between all those that fall within a pre-assigned cutoff distance. We considered this distance to be 5.64 Å, four times the equilibrium C–N bond length [42, 43]. After the new bonds were identified, all new additional covalent terms were created and hydrogen atoms from the reactive C and N atoms were removed. Then several 50 ps NPT simulations were preformed until no reactive pairs exist within the cut-off distance. At the end, the structure was again equilibrated for 200 ps in the NVT ensemble at 300 K.

After the energy minimization process, the MD cell was strained in both tension and compression to determine the bulk moduli applying equal strains in the loading directions along all three axes; while the average shear modulus was determined by applying equal shear strains on the simulation box in xy, xz, and yz planes. In all simulations, strain increments of 0.25% were applied along a particular direction by uniformly deforming or shearing the MD simulation box and updating the atom coordinates to fit within the new cell dimensions. After each strain increment, the MD unit cell was equilibrated in the NVT ensemble for 10 ps at 300 K. It may be noted that the fluctuations in the temperature and potential energy profiles are less than 1% when the system reached equilibrium after about 5 ps [44] and several existing MD studies [34, 44-46] used 2 ps to 10 ps time step in their MD simulations to equilibrate the systems after each strain increment. Then, the stress tensor was averaged over an interval of 10 ps to reduce the fluctuation effects. These steps were repeated again in the subsequent strain increments until the total strain reached up to 2.5%. Based on the calculated bulk and shear moduli, Young's modulus (E) and Poisson's ratio (v) were determined. The predicted elastic properties of the epoxy using MD simulations are summarized in Table 2 and Young's modulus was found to be consistent with the experimentally measured modulus of a similar epoxy [47].

### 2.1.2. MD simulations of nano-fiber

The noncovalent bonded nano-fiber made of CNT-epoxy nanocomposite system is considered herein; therefore, the interactions between the atoms of the embedded CNT and the surrounding epoxy are solely from non-bonded interactions: van der Waals (vdW) interactions and Coulombic forces. We considered the cut-off distance as 14.0 Å for the non-bonded interaction [44]. At the vicinity of CNT-matrix interface, the structure of the epoxy matrix differs from the bulk epoxy and an ultra-thin epoxy layer formed due to chemical reactions. This ultra-thin layer consists of highly packed crystalline epoxy monomers, which has different elastic properties in comparison to the amorphous bulk polymer [48, 49]. In order to obtain the actual nano-fiber properties, the size of the MD unit cell must be large enough to incorporate the change of the polymer structure. The cylindrical molecular structure of the CNT was considered as an equivalent cylindrical fiber [46, 50, 51] to determine its volume fraction in the nano-fiber,

$$\text{CNT volume fraction} = \frac{L_{CNT} \times \frac{\pi}{4} \left[ (d_{CNT} + h_{vdw})^2 - (d_{CNT} - t_{CNT})^2 \right]}{V} \tag{2}$$

where $d_{CNT}$, $L_{CNT}$ and $t_{CNT}$ are the diameter, length and thickness of a CNT, respectively, and $h_{vdW}$ is vdW equilibrium distance between the CNT and the surrounding polymer matrix.

We performed a MD simulation for a system consisting of a single CNT of length 43.0 Å embedded in 51 epoxy oligomers to determine the thickness of the interface layer. The initial size of the periodic MD cell was 150 Å × 150 Å × 150 Å. Subsequently, the volume of the simulation box was gradually reduced to 50 Å × 50 Å × 43 Å and equilibrated using the same steps as adopted in the pure epoxy case (see Fig. 4). The details of reduced unit cell are summarized in Table 1. The developed epoxy layer at the CNT and bulk matrix interface exists due to the nonbonded interactions. Figure 5 shows the radial distribution function (RDF) of the epoxy atoms surrounding the CNT in the equilibrium state. The variation of the RDF along the radial direction represents the changes in the epoxy structure at the vicinity of CNT. Figure 5 demonstrates that the RDF is zero at the radial distance of 8 Å and increases at its maximum value at the radial distance of 9.5 Å. Thereafter, it starts to fluctuate around an average value of 220 atoms/nm$^3$. Subsequently, the appropriate MD unit cell size for nano-fiber was selected using the determined values of the equilibrium separation distance and the interface thickness. The MD unit cell for nano-fiber is assumed to be transversely isotropic with the 3–axis being the axis of symmetry; therefore, only five independent elastic coefficients are required to define the elastic stiffness matrix. The stress tensor for the transversely isotropic MD unit cell is:

$$\begin{bmatrix} \sigma_{11} \\ \sigma_{22} \\ \sigma_{33} \\ \sigma_{23} \\ \sigma_{13} \\ \sigma_{12} \end{bmatrix} = \begin{bmatrix} C_{11} & C_{12} & C_{13} & 0 & 0 & 0 \\ C_{12} & C_{11} & C_{13} & 0 & 0 & 0 \\ C_{13} & C_{13} & C_{33} & 0 & 0 & 0 \\ 0 & 0 & 0 & C_{44} & 0 & 0 \\ 0 & 0 & 0 & 0 & C_{44} & 0 \\ 0 & 0 & 0 & 0 & 0 & (C_{11} - C_{12})/2 \end{bmatrix} \begin{bmatrix} \varepsilon_{11} \\ \varepsilon_{22} \\ \varepsilon_{33} \\ \varepsilon_{23} \\ \varepsilon_{13} \\ \varepsilon_{12} \end{bmatrix} \qquad (3)$$

where $\sigma_{ij}$ and $\varepsilon_{ij}$ are the stress and strain components, respectively, and $C_{ij}$ represents the elastic coefficients of the unit cell. A series of MD simulations were carried out to determine the elastic moduli of the unit cell. The MD unit cell was constructed by placing the crosslinked epoxy chains around the CNT. Five sets of boundary conditions were chosen to determine five independent elastic constants such that a single property can be independently determined for each boundary condition. The displacements applied at the boundary of the unit cell are summarized in Table 3; in which symbols have usual meaning. The steps involved in the MD simulations of the unit cell of nano-fiber are the same as adopted in the case of pure epoxy. Table 4 summarizes the outcome of the MD simulations. These properties of the unit cell will be used as the properties of nano-fiber in the micromechanical model to determine the effective elastic moduli of the CNRP at the microscale level (see Fig. 1).

## 2.2. Effective elastic properties of CNRP

In this section, the elastic properties of the pure epoxy and the nano-fiber obtained from the MD simulations were used as input in the Mori-Tanaka model in order to determine the effective elastic properties of the CNRP. The augmented carbon fiber in the CNRP can be viewed as a composite fiber, as shown in Fig. 6. The intermediate phase was considered to be made of CNRP. Practically, the orientations of the CNT reinforcement in the polymer matrix vary over the nanocomposite volume. Therefore, studying the properties of CNRP reinforced with aligned CNTs is of a great importance, especially randomly oriented CNTs. Here, we consider three different cases of CNRP, in which nano-fibers are: (i) aligned along the direction of fiber, (ii) aligned radially to the axis of fiber, and (iii) randomly dispersed.

Mori-Tanaka model [52] is an Eshelby-type model which accounts for the interactions among the neighboring reinforcements. Due to its simplicity, the Mori-Tanaka model has been reported to be the efficient analytical micromechanical model for estimating the effective orthotropic elastic properties of the composites. Hence, Mori-Tanaka model is used to determine the effective orthotropic elastic properties of the CNRP which are required as input to the shear lag model development presented in Section 3. From the constructional feature of the nano-reinforced composite, it may be viewed that the carbon fiber is embedded in the CNRP material. Employing the procedure of the Mori-Tanaka model for the cylindrical inclusions [53], the effective elastic properties of CNRP can be estimated using the following explicit formulation:

$$[C] = [C^{nf}] + v_{nf}([C^{nf}] - [C^m])\left([\tilde{A}_1]\left[v_m[I] + v_{nf}[\tilde{A}_1]\right]^{-1}\right) \qquad (4a)$$

in which

$$[\tilde{A}_1] = \left[[I] + [S^E]([C^m])^{-1}([C^{nf}] - [C^m])\right]^{-1}$$

where $[C^{nf}]$ and $[C^m]$ are the stiffness tensors of the nano-fiber and the epoxy matrix, respectively; $[I]$ is an identity tensor; $v_{nf}$ and $v_m$ represent the volume fractions of the nano-fiber and the epoxy matrix, respectively; and $[S^E]$ indicates the Eshelby tensor and its elements are explicitly given in Appendix A.

In the cases (ii) and (iii) of epoxy reinforced with radially and randomly oriented nano-fibers in the three-dimensional space, respectively, the following relation can be used to determine the effective elastic properties of the CNRP:

$$[C] = [C^m] + v_{nf}([C^{nf}] - [C^m])\left([\tilde{A}_2]\left[v_m[I] + v_{nf}[\langle[\tilde{A}_2]\rangle]\right]^{-1}\right) \qquad (4b)$$

in which

$$[\tilde{A}_2] = \left[[I] + [S^E]([C^m])^{-1}([C^{nf}] - [C^m])\right]^{-1}$$

The terms enclosed with angle brackets in Eq. (4b) represent the average value of the term over all orientations defined by local-global coordinate transformation. The strain concentration tensor for the nano-fiber with respect to the global coordinates is given by

$$[\tilde{A}_{ijkl}] = t_{ip}t_{jq}t_{kr}t_{ls}[A_{pqrs}] \qquad (5)$$

where $t_{ij}$ are the direction cosines for the transformation and are given in Appendix A.

Finally, the random orientation average of the strain concentration tensor $\langle[\tilde{A}_2]\rangle$ can be determined by using the following relation (Marzari and Ferrari, 1992):

$$\langle[\tilde{A}_2]\rangle = \frac{\int_{-\pi}^{\pi}\int_0^{\pi}\int_0^{\pi/2}[\tilde{A}](\phi,\gamma,\psi)\,\sin\gamma\,d\phi d\gamma d\psi}{\int_{-\pi}^{\pi}\int_0^{\pi}\int_0^{\pi/2}\sin\gamma\,d\phi d\gamma d\psi} \qquad (6)$$

where $\phi$, $\gamma$, and $\psi$ are the Euler angles with respect to 1, 2, and 3 axes..

## 3. Shear lag model formulation

Figure 7 represents the in-plane cross section of the nano-reinforced composite lamina in which the composite fibers are packed in hexagonal array in the transverse plane. The composite fibers are assumed to be uniformly dispersed over the volume of the composite lamina in such a way that the three orthogonal principal material coordinate axes exist in it. This lamina can be viewed to be composed of either the RVEs I or the RVEs II, arranged in a periodic manner in the polymer matrix. Based on the cross-section of the cylindrical RVE I shown in Fig. 8, an analytical three-phase shear lag model derived herein. It may be noted that many pioneering researchers [2-5] considered such cylindrical RVE in their two- and three-phase shear lag studies. The nano-reinforced composite and its constituent phases are being studied here are made of aligned reinforcements. Therefore, a three-phase shear lag model was developed in this Section considering the orthotropy of the nano-reinforced composite and its constituent phases. In our model, the application of the shear stress ($\tau$) along the length of the RVE at r = R [54-56] accounts for the staggering of the adjacent RVEs, whereas the consideration of the radial load (q) on the RVE accounts for the lateral extensional interaction between the adjacent RVEs. Such radial/residual stresses may arise due to the manufacturing processes such as injection molding of the short fiber composites.

The cylindrical coordinate system (r–θ–x) is considered in such a way that the RVE axis coincides with the x–axis. The model was derived by dividing the RVE into three zones. In the RVE zone $-L_f \leq x \leq L_f$ consists of three concentric cylindrical phases; namely, the carbon fiber, the CNRP and the polymer matrix. The RVE of the nano-reinforced composite has the radius R and the length 2L while a and $2L_f$ denote the radius and the length of carbon fiber, respectively. The inner and outer radii of the CNRP phase are a and b, respectively. The respective portions of the RVE in the zones $-L \leq x \leq -L_f$ and $L_f \leq x \leq L$ are considered to be composed of an imaginary fiber, an imaginary CNRP and the polymer matrix phase. The geometric dimensions of imaginary carbon fiber and CNRP are same as those of actual ones. Thus, the shear lag model derived herein for the zone $-L_f \leq x \leq L_f$ can be applied to derive the shear lag models for the zones $-L \leq x \leq -L_f$ and $L_f \leq x \leq L$. In what follows, the shear lag model for the zone $-L_f \leq x \leq L_f$ is first derived.

## 3.1. Governing equations of stress transfer analysis for the middle portion $(-L_f \leq x \leq L_f)$

The governing equilibrium equations for an axisymmetric RVE problem in terms of the cylindrical coordinates (r–θ–x) are given by (Timoshenko and Goodier, 1970):

$$\frac{\partial \sigma_r^k}{\partial r} + \frac{\partial \sigma_{xr}^k}{\partial x} + \frac{\sigma_r^k - \sigma_\theta^k}{r} = 0 \quad \text{and} \quad \frac{\partial \sigma_x^k}{\partial x} + \frac{1}{r}\frac{\partial (r\sigma_{xr}^k)}{\partial r} = 0 ; \quad k = f, c \text{ and } m \quad (7a, b)$$

while the relevant constitutive relations are

$$\sigma_x^k = C_{11}^k \epsilon_x^k + C_{12}^k \epsilon_\theta^k + C_{13}^k \epsilon_r^k, \qquad \sigma_r^k = C_{13}^k \epsilon_x^k + C_{23}^k \epsilon_\theta^k + C_{33}^k \epsilon_r^k \text{ and}$$

$$\sigma_{xr}^k = C_{66}^k \epsilon_{xr}^k ; \quad k = f, c \text{ and } m \quad (8)$$

In Eqs. (7) and (8), the respective superscripts f, c and m denote the carbon fiber, the CNRP and the polymer matrix. For the $k^{th}$ constituent phase, $\sigma_x^k$ and $\sigma_r^k$ represent the normal stresses in the x and r directions, respectively; $\epsilon_x^k$, $\epsilon_\theta^k$ and $\epsilon_r^k$ are the normal strains along the x, θ and r, directions, respectively; $\sigma_{xr}^k$ is the transverse shear stress, $\epsilon_{xr}^k$ is the transverse shear strain and $C_{ij}^k$ are the elastic constants. The strain-displacement relations for an axisymmetric problem relevant to this RVE are

$$\epsilon_x^k = \frac{\partial u^k}{\partial x}, \epsilon_\theta^k = \frac{w^k}{r}, \epsilon_r^k = \frac{\partial w^k}{\partial r} \text{ and } \epsilon_{xr}^k = \frac{\partial u^k}{\partial r} + \frac{\partial w^k}{\partial x} \quad (9)$$

where $u^k$ and $w^k$ denote the axial and radial displacements at any point of the $k^{th}$ phase along the x and r directions, respectively. We can write the following traction boundary conditions

$$\sigma_r^m|_{r=R} = q \quad \text{and} \quad \sigma_{xr}^m|_{r=R} = \tau = \frac{2R\sigma}{Lv_{CF}} \quad (10)$$

and the interfacial traction continuity conditions are given by

$$\sigma_r^f|_{r=a,-L_f \leq x \leq L_f} = \sigma_r^c|_{r=a,-L_f \leq x \leq L_f} ; \quad \sigma_{xr}^f|_{r=a,-L_f \leq x \leq L_f} = \sigma_{xr}^c|_{r=a,-L_f \leq x \leq L_f} = \tau_1 ;$$

$$\sigma_r^c|_{r=b,-L_f \leq x \leq L_f} = \sigma_r^m|_{r=b,-L_f \leq x \leq L_f}; \quad \sigma_{xr}^c|_{r=b,-L_f \leq x \leq L_f} = \sigma_{xr}^m|_{r=b,-L_f \leq x \leq L_f} = \tau_2 ;$$

$$u^f|_{r=a,-L_f \leq x \leq L_f} = u^c|_{r=a,-L_f \leq x \leq L_f}; \quad u^c|_{r=b,-L_f \leq x \leq L_f} = u^m|_{r=b,-L_f \leq x \leq L_f};$$

$$w^f|_{r=a,-L_f \leq x \leq L_f} = w^c|_{r=a,-L_f \leq x \leq L_f} \text{ and } w^c|_{r=b,-L_f \leq x \leq L_f} = w^m|_{r=b,-L_f \leq x \leq L_f} \quad (11)$$

where $\tau_1$ is the transverse shear stress at the interface between the carbon fiber and the CNRP while $\tau_2$ is the transverse shear stress at the interface between the CNRP and the polymer.

The average axial stresses in the different phases are defined as

$$\overline{\sigma}_x^f = \frac{1}{\pi a^2} \int_0^a \sigma_x^f 2\pi r \, dr \, ;$$

$$\overline{\sigma}_x^c = \frac{1}{\pi(b^2 - a^2)} \int_a^b \sigma_x^c 2\pi r \, dr \quad \text{and} \quad \overline{\sigma}_x^m = \frac{1}{\pi(R^2 - b^2)} \int_b^R \sigma_x^m 2\pi r \, dr \tag{12}$$

Now, making use of Eqs. (7) and (10–12), it can be shown that

$$\frac{\partial \overline{\sigma}_x^f}{\partial x} = -\frac{2}{a} \tau_1 \, ; \, \frac{\partial \overline{\sigma}_x^c}{\partial x} = \frac{2a}{b^2 - a^2} \tau_1 - \frac{2b}{b^2 - a^2} \tau_2 \text{ and } \frac{\partial \overline{\sigma}_x^m}{\partial x} = \frac{2b}{R^2 - b^2} \tau_2 \tag{13}$$

It is evident from Eq. (13) that the gradients of $\overline{\sigma}_x^c$ and $\overline{\sigma}_x^m$ with respect to the axial coordinate (x) are independent of the radial coordinate (r). Hence, as the radial dimension of the RVE is very small, it is reasonable to assume that [2]

$$\frac{\partial \sigma_x^k}{\partial x} = \frac{\partial \overline{\sigma}_x^k}{\partial x} \, ; \, c \text{ and } m \tag{14}$$

Thus, using the equilibrium equation given by Eq. (7b), the transverse shear stresses in the CNRP phase and the polymer matrix phase can be expressed in terms of the interfacial shear stresses $\tau_1$ and $\tau_2$, respectively, as follows:

$$\sigma_{xr}^c = \frac{a}{r} \tau_1 + \frac{1}{2r}(a^2 - r^2) \frac{\partial \overline{\sigma}_x^c}{\partial x} \tag{15}$$

$$\sigma_{xr}^m = \left(\frac{R^2}{r} - r\right) \frac{b}{R^2 - b^2} \tau_2 + \frac{R}{r} \tau \tag{16}$$

Also, since the RVE is an axisymmetric problem, it is usually assumed [2, 57] that the gradient of the radial displacements with respect to the x–direction is negligible and so, from the constitutive relation given by Eq. (8) and the strain-displacement relations given by Eq. (9) between $\sigma_{xr}^k$ and $\epsilon_{xr}^k$, one can write

$$\frac{\partial u^k}{\partial r} \approx \frac{1}{C_{66}^k} \sigma_{xr}^k \, ; \, c \text{ and } m \tag{17}$$

Solving Eq. (17) and satisfying the continuity conditions at r = a and r = b, respectively, the axial displacements of the CNRP phase and the polymer matrix phase along the x–direction can be derived as follows:

$$u^c = u_a^f + A_1 \tau_1 + A_2 \tau_2 \tag{18}$$

$$u^m = u_a^f + A_3 \tau_1 + A_4 \tau_2 + \frac{R}{C_{66}^m} \tau \ln\left(\frac{r}{b}\right) \text{ and} \tag{19}$$

$$u_a^f = u^f\big|_{r=a} \tag{20}$$

in which $A_i$ (i = 1, 2, 3 and 4) are the constants of the displacement fields of the CNRP and the polymer matrix, and are explicitly shown in Appendix C.

The radial displacements in the three constituent phases can be assumed as [58]

$$w^f = C_1 r, \quad w^c = A_c r + \frac{B_c}{r} \text{ and } w^m = C_2 r + \frac{C_3}{r} \tag{21}$$

where $C_1, A_c, B_c, C_2$ and $C_3$ are the unknown constants. Invoking the continuity conditions for the radial displacement at the interfaces r = a and b, the radial displacement in the CNRP phase can be augmented as follows:

$$w^c = \frac{a^2}{b^2 - a^2}\left(\frac{b^2}{r} - r\right)C_1 - \frac{b^2}{b^2 - a^2}\left(\frac{a^2}{r} - r\right)C_2 - \frac{1}{b^2 - a^2}\left(\frac{a^2}{r} - r\right)C_3 \tag{22}$$

Substituting Eqs. (18), (19), (21) and (22) into Eq. (9) and subsequently, employing the constitutive relations (8), the expressions for the normal stresses can be written in terms of the unknown constants $C_1, C_2$ and $C_3$ as follows:

$$\overline{\sigma}_x^f = C_{11}^f u_a^{f\prime} + 2C_{12}^f C_1 \tag{23}$$

$$\sigma_r^f = \frac{C_{12}^f}{C_{11}^f}\overline{\sigma}_x^f + \left[C_{23}^f + C_{33}^f - \frac{2\left(C_{12}^f\right)^2}{C_{11}^f}\right]C_1 \tag{24}$$

$$\sigma_x^c = \frac{C_{11}^c}{C_{11}^f}\overline{\sigma}_x^f - \left(\frac{2C_{12}^c a^2}{b^2 - a^2} + \frac{2C_{12}^f C_{11}^c}{C_{11}^f}\right)C_1 + \frac{2C_{12}^c b^2}{b^2 - a^2}C_2 + \frac{2C_{12}^c}{b^2 - a^2}C_3 + C_{11}^c A_1 \tau_1'$$
$$+ C_{11}^c A_2 \tau_2' \tag{25}$$

$$\sigma_r^c = \frac{C_{13}^c}{C_{11}^f}\overline{\sigma}_x^f + \left[\frac{C_{13}^c a^2}{b^2 - a^2}\left(\frac{b^2}{r^2} - 1\right) + \frac{C_{33}^c a^2}{b^2 - a^2}\left(-\frac{b^2}{r^2} - 1\right) - \frac{2C_{12}^f C_{13}^c}{C_{11}^f}\right]C_1$$

$$+\left[-\frac{C_{13}^c b^2}{b^2-a^2}\left(\frac{a^2}{r^2}-1\right)+\frac{C_{33}^c b^2}{b^2-a^2}\left(\frac{a^2}{r^2}+1\right)\right]C_2$$

$$+\left[-\frac{C_{23}^c}{b^2-a^2}\left(\frac{a^2}{r^2}-1\right)+\frac{C_{33}^c}{b^2-a^2}\left(\frac{a^2}{r^2}+1\right)\right]C_3+C_{13}^c A_1\tau_1'+C_{13}^c A_2\tau_2' \qquad (26)$$

$$\sigma_x^m=\frac{C_{11}^m}{C_{11}^f}\overline{\sigma}_x^f-\frac{2C_{12}^m C_{11}^m}{C_{11}^f}C_1+2C_{12}^m C_2+C_{11}^m A_3\tau_1'+C_{11}^m A_4\tau_2'+\frac{R}{C_{66}^m}\tau\ln\left(\frac{r}{b}\right)C_{11}^m \qquad (27)$$

$$\sigma_r^m=\frac{C_{12}^m}{C_{11}^f}\overline{\sigma}_x^f-\frac{2C_{12}^f C_{12}^m}{C_{11}^f}C_1+(C_{11}^m+C_{12}^m)C_2+(C_{12}^m-C_{11}^m)\frac{C_3}{r^2}+C_{12}^m A_3\tau_1'$$

$$+C_{12}^m A_4\tau_2'+\frac{R}{C_{66}^m}\tau\ln\left(\frac{r}{b}\right)C_{12}^m \qquad (28)$$

where the prime notations denote the differentiation with respect to the axial coordinate (x). Invoking the continuity conditions $\sigma_r^f\big|_{r=a}=\sigma_r^c\big|_{r=a}$ and $\sigma_r^c\big|_{r=b}=\sigma_r^m\big|_{r=b}$, and satisfying the boundary condition $\sigma_r^m\big|_{r=R}=q$, the following equations for solving $C_1$, $C_2$ and $C_3$ are obtained:

$$\begin{bmatrix}B_{11}&B_{12}&B_{13}\\B_{21}&B_{22}&B_{23}\\B_{31}&B_{32}&B_{33}\end{bmatrix}\begin{Bmatrix}C_1\\C_2\\C_3\end{Bmatrix}=\frac{\overline{\sigma}_x^f}{C_{11}^f}\begin{Bmatrix}C_{12}^f-C_{13}^c\\C_{13}^c-C_{12}^m\\-C_{12}^m\end{Bmatrix}+\begin{Bmatrix}0\\C_{13}^c A_5-C_{12}^m A_3\\-C_{12}^m A_3\end{Bmatrix}\tau_1'$$

$$+\begin{Bmatrix}0\\C_{12}^m A_7-C_{13}^c A_6\\-C_{12}^m A_8\end{Bmatrix}\tau_2'+\begin{Bmatrix}0\\0\\1\end{Bmatrix}q+\begin{Bmatrix}0\\0\\-\frac{R}{C_{66}^m}\ln\left(\frac{R}{b}\right)C_{12}^m\end{Bmatrix}\tau \qquad (29)$$

The expressions of the coefficients $B_{ij}$ are presented in Appendix C. Solving Eq. (29), the solutions of the constants $C_1$, $C_2$ and $C_3$ can be expressed as:

$$C_i=b_{i1}\overline{\sigma}_x^f+b_{i2}\tau_1'+b_{i3}\tau_2'+b_{i4}q+b_{i5}\tau; \quad i=1,2,3 \qquad (30)$$

The expressions of the coefficients $b_{i1}$, $b_{i2}$, $b_{i3}$, $b_{i4}$ and $b_{i5}$ are evident from Eq. (30), and are not shown here for the sake of clarity. Now, making use of Eqs. (25), (27) and (30) in the last two equations of (12), respectively, the average axial stresses in the CNRP phase and the polymer matrix phase are written as follows:

$$\overline{\sigma}_x^c=A_{14}\overline{\sigma}_x^f+A_{15}\tau_1'+A_{16}\tau_2'+A_{17}q+A_{18}\tau \qquad (31)$$

$$\overline{\sigma}_x^m=A_{20}\overline{\sigma}_x^f+A_{21}\tau_1'+A_{22}\tau_2'+A_{23}q+A_{24}\tau \qquad (32)$$

The constants $A_i$ ($i = 5, 6, 7, \ldots., 24$) appeared in the above four equations are presented in Appendix C. Now, satisfying the equilibrium of stress along the axial direction at any transverse cross-section of the RVE I, the following equation was obtained:

$$\pi R^2 \sigma = \pi a^2 \overline{\sigma}_x^f + \pi(b^2 - a^2)\overline{\sigma}_x^c + \pi(R^2 - b^2)\overline{\sigma}_x^m \qquad (33)$$

Differentiating the first and last equations of (13) with respect to x, we have

$$\tau_1' = -\frac{a}{2}\overline{\sigma}_x^{f\,''} \text{ and } \tau_2' = \frac{R^2 - b^2}{2b}\overline{\sigma}_x^{m\,''} \qquad (34)$$

Use of Eqs. (31 –34), yields

$$A_{25}\overline{\sigma}_x^f + A_{26}\overline{\sigma}_x^{f\,''} + A_{27}\overline{\sigma}_x^{m\,''} + A_{28}q + A_{29}\tau - R^2\sigma = 0 \qquad (35)$$

Deriving the expression for $\tau_2'$ from Eq. (31) and substituting the same into Eq. (32), and then using Eq. (34), the following result for $\overline{\sigma}_x^m$ was obtained:

$$\overline{\sigma}_x^m = \left(A_{20} - \frac{A_{14}A_{22}}{A_{16}}\right)\overline{\sigma}_x^f + \left(\frac{A_{22}}{A_{16}}\right)\overline{\sigma}_x^c + \left(\frac{a}{2}\right)\left(\frac{A_{22}A_{15}}{A_{16}} - A_{21}\right)\overline{\sigma}_x^{f\,''} + \left(A_{23} - \frac{A_{17}A_{22}}{A_{16}}\right)q$$

$$+ \left(A_{24} - \frac{A_{18}A_{22}}{A_{16}}\right)\tau \qquad (36)$$

Differentiating Eqs. (33) and (36) twice with respect to x and using the resulting equations in Eq. (35), the governing equation for the average axial stress in the carbon fiber was obtained as follows:

$$\overline{\sigma}_x^{f\,(4)} + A_{31}\overline{\sigma}_x^{f\,''} + A_{32}\overline{\sigma}_x^f - A_{33}\sigma + A_{34}q + A_{35}\tau = 0 \qquad (37)$$

The constants $A_i$ ($i = 25, 26, 27, \ldots., 35$) appeared in the above three equations are presented in Appendix C. Solution of Eq. (37) is given by:

$$\overline{\sigma}_x^f = A_{36}\sinh(\alpha x) + A_{37}\cosh(\alpha x) + A_{38}\sinh(\beta x) + A_{39}\cosh(\beta x) + (A_{33}/A_{32})\sigma$$

$$- (A_{34}/A_{32})q - (A_{35}/A_{32})\tau \qquad (38)$$

where

$$\alpha = \sqrt{1/2\left(-A_{31} + \sqrt{(A_{31})^2 - 4A_{32}}\right)} \text{ and } \beta = \sqrt{1/2\left(-A_{31} - \sqrt{(A_{31})^2 - 4A_{32}}\right)} \qquad (39)$$

Substitution of Eq. (39) into the first equation of (13) yields the expression for the carbon fiber/CNRP interfacial shear stress as follows:

$$\tau_1 = -\frac{a}{2}\left[A_{36}\alpha\cosh(\alpha x) + A_{37}\alpha\sinh(\alpha x) + A_{38}\beta\cosh(\beta x) + A_{39}\beta\sinh(\beta x)\right] \quad (40)$$

### 3.2. Governing equations of stress transfer analysis for the two end portions $(-L \leq x \leq -L_f$ and $L_f \leq x \leq L)$

The two-end portions of the RVE I in the zones $-L \leq x \leq -L_f$ and $L_f \leq x \leq L$ are considered to be comprised of imaginary carbon fiber (pf) and CNRP (pc) made of the polymer material. The radius of the imaginary fiber is denoted by a while the inner and outer radii of the imaginary CNRP phase are represented by a and b, respectively. Thus, the solutions derived for the middle portion of the RVE I $(-L_f \leq x \leq L_f)$ can be applied to derive the solutions for the zones $-L \leq x \leq -L_f$ and $L_f \leq x \leq L$ by substituting $C_{ij}^f = C_{ij}^c = C_{ij}^m$. The traction boundary conditions and the interfacial continuity conditions are given by

$$\overline{\sigma}_x^{pf}\Big|_{x=\pm L, 0 \leq r \leq a} = \sigma \quad \text{and} \quad \tau_1^{pf}\Big|_{r=a} = 0 \text{ at } x = \pm L \quad (41a)$$

$$\overline{\sigma}_x^f\Big|_{x=\pm L_f, 0 \leq r \leq a} = \overline{\sigma}_x^{pf}\Big|_{x=\pm L_f, 0 \leq r \leq a} \quad \text{and} \quad \tau_1|_{r=a,} = \tau_1^{pf}\Big|_{r=a} \text{ at } x = \pm L_f \quad (41b)$$

$$\overline{\sigma}_x^{pc}\Big|_{x=\pm L, a \leq r \leq b} = \sigma \quad \text{and} \quad \tau_2^{pc}\Big|_{r=b} = 0 \text{ at } x = \pm L \quad (42a)$$

$$\overline{\sigma}_x^c\Big|_{x=\pm L_f, a \leq r \leq b} = \overline{\sigma}_x^{pc}\Big|_{x=\pm L_f, a \leq r \leq b} \quad \text{and} \quad \tau_2^c\big|_{r=b,} = \tau_2^{pc}\Big|_{r=b} \text{ at } x = \pm L_f \quad (42b)$$

$$\sigma_r^{pf}\Big|_{r=a, -L \leq x \leq -L_f, L_f \leq x \leq L} = \sigma_r^{pc}\Big|_{r=a, -L \leq x \leq -L_f, L_f \leq x \leq L} \; ;$$

$$\sigma_{xr}^{pf}\Big|_{r=a, -L \leq x \leq -L_f, L_f \leq x \leq L} = \sigma_{xr}^{pc}\Big|_{r=a, -L \leq x \leq -L_f, L_f \leq x \leq L} = \tau_1^{pf} \; ;$$

$$\sigma_r^{pc}\Big|_{r=b, -L \leq x \leq -L_f, L_f \leq x \leq L} = \sigma_r^m\Big|_{r=b, -L \leq x \leq -L_f, L_f \leq x \leq L} ;$$

$$\sigma_{xr}^{pc}\Big|_{r=b, -L \leq x \leq -L_f, L_f \leq x \leq L} = \sigma_{xr}^m\Big|_{r=b, -L \leq x \leq -L_f, L_f \leq x \leq L} = \tau_2^{pc} \; ;$$

$$u^{pf}\Big|_{r=a, -L \leq x \leq -L_f, L_f \leq x \leq L} = u^{pc}\Big|_{r=a, -L \leq x \leq -L_f, L_f \leq x \leq L} ;$$

$$u^{pc}\Big|_{r=b, -L \leq x \leq -L_f, L_f \leq x \leq L} = u^m\Big|_{r=b, -L \leq x \leq -L_f, L_f \leq x \leq L} ;$$

$$w^{pf}\big|_{r=a,-L\leq x\leq -L_f, L_f\leq x\leq L} = w^{pc}\big|_{r=a,-L\leq x\leq -L_f, L_f\leq x\leq L} \quad \text{and}$$

$$w^{pc}\big|_{r=b,-L\leq x\leq -L_f, L_f\leq x\leq L} = w^{m}\big|_{r=b,-L\leq x\leq -L_f, L_f\leq x\leq L} \tag{43}$$

where the superscript pf and pc represent imaginary carbon fiber and CNRP material made of the polymer material, respectively. Following the solution procedure adopted in Section 4.1, the governing equation for the average axial stress ($\overline{\sigma}_x^{pf}$) in an imaginary fiber made of the polymer material lying in the zones $-L \leq x \leq -L_f$ and $L_f \leq x \leq L$ can be written as

$$\overline{\sigma}_x^{pf^{(4)}} + A_{31}^{pf}\overline{\sigma}_x^{pf''} + A_{32}^{pf}\overline{\sigma}_x^{pf} - A_{33}^{pf}\sigma + A_{34}^{pf}q + A_{35}^{pf}\tau = 0 \tag{44}$$

In the above equation, the expressions for the constants $A_{31}^{pf}-A_{35}^{pf}$ are similar to those of the expressions $A_{31}-A_{35}$, respectively, but these are to be derived by considering $C_{ij}^f = C_{ij}^c = C_{ij}^m$. Solution of Eq. (44) is given by:

$$\overline{\sigma}_x^{pf} = \left(A_{33}^{pf}/A_{32}^{pf}\right)\sigma - \left(A_{34}^{pf}/A_{32}^{pf}\right)q - \left(A_{35}^{pf}/A_{32}^{pf}\right)\tau + A_{36}^{pf}\sinh(\alpha^{pf}x) + A_{37}^{pf}\cosh(\alpha^{pf}x)$$

$$+ A_{38}^{pf}\sinh(\beta^{pf}x) + A_{39}^{pf}\cosh(\beta^{pf}x) \tag{45}$$

$$\alpha^{pf} = \sqrt{1/2\left(-A_{31}^{pf} + \sqrt{\left(A_{31}^{pf}\right)^2 - 4A_{32}^{pf}}\right)} \quad \text{and} \quad \beta^{pf} = \sqrt{1/2\left(-A_{31}^{pf} - \sqrt{\left(A_{31}^{pf}\right)^2 - 4A_{32}^{pf}}\right)}$$

Utilizing the end conditions given by Eq. (41a) in Eq. (45), the constants $A_{36}^{pf}$, $A_{37}^{pf}$, $A_{38}^{pf}$ and $A_{39}^{pf}$ can be explicitly written as follows:

$$A_{36}^{pf} = 0 \tag{46}$$

$$A_{37}^{pf} = \frac{\beta^{pf}\sinh(\beta^{pf}L)}{\beta^{pf}\sinh(\beta^{pf}L)\cosh(\alpha^{pf}L) - \alpha^{pf}\sinh(\alpha^{pf}L)\cosh(\beta^{pf}L)}$$

$$\times \left\{\sigma - \frac{A_{33}^{pf}}{A_{32}^{pf}}\sigma + \frac{A_{34}^{pf}}{A_{32}^{pf}}q + \frac{A_{35}^{pf}}{A_{32}^{pf}}\tau\right\} \tag{47}$$

$$A_{38}^{pf} = 0 \tag{48}$$

$$A_{39}^{pf} = -\frac{\alpha^{pf}\sinh(\alpha^{pf}L)}{\beta^{pf}\sinh(\beta^{pf}L)\cosh(\alpha^{pf}L) - \alpha^{pf}\sinh(\alpha^{pf}L)\cosh(\beta^{pf}L)}$$

$$\times \left\{ \sigma - \frac{A_{33}^{pf}}{A_{32}^{pf}}\sigma + \frac{A_{34}^{pf}}{A_{32}^{pf}}q + \frac{A_{35}^{pf}}{A_{32}^{pf}}\tau \right\} \tag{49}$$

Substituting Eqs. (46–49) in Eq. (45), the final solution for $\overline{\sigma}_x^{pf}$ was obtained as follows:

$$\overline{\sigma}_x^{pf} = \left[ \frac{\beta^{pf}\sinh(\beta^{pf}L)\cosh(\alpha^{pf}x) - \alpha^{pf}\sinh(\alpha^{pf}L)\cosh(\beta^{pf}x)}{\beta^{pf}\sinh(\beta^{pf}L)\cosh(\alpha^{pf}L) - \alpha^{pf}\sinh(\alpha^{pf}L)\cosh(\beta^{pf}L)} \right]$$

$$\times \left\{ \sigma - \frac{A_{33}^{pf}}{A_{32}^{pf}}\sigma + \frac{A_{34}^{pf}}{A_{32}^{pf}}q + \frac{A_{35}^{pf}}{A_{32}^{pf}}\tau \right\} + \frac{A_{33}^{pf}}{A_{32}^{pf}}\sigma - \frac{A_{34}^{pf}}{A_{32}^{pf}}q - \frac{A_{35}^{pf}}{A_{32}^{pf}}\tau \tag{50}$$

Similarly, utilizing Eq. (45) and the end conditions given by Eq. (41b) in Eq. (38), the constants $A_{36}$, $A_{37}$, $A_{38}$ and $A_{39}$ are evaluated to determine $\overline{\sigma}_x^f$.

## 4. Results and Discussion

In this Section, the predictions by the Mori-Tanaka model were first compared with those of the existing numerical results of the CNT-reinforced composite to verify the validity of the CNRP modeling. Subsequently, the results from the analytical shear lag model were compared with the FE shear lag simulations and were used to analyze the stress transfer characteristics of the nano-reinforced composite. Carbon fiber and polyimide matrix were used for evaluating the numerical results. Their material properties, available in References [59, 60] are listed in Table 5. Unless otherwise mentioned, the following geometrical parameters of the RVE of the nano-reinforced composite are adopted to compute the results in the present study: carbon fiber volume fraction ($v_f$) = 0.4; composite fiber volume fraction in the RVE ($v_{CF}$) = 0.7513; carbon fiber radius (a) = 5 μm; composite fiber radius (b) = 6.5258 μm; RVE radius (R) = 7.1783 μm; half length of the carbon fiber ($L_f$) = 100 μm; half length of the RVE (L) = 110 μm; width of the RVE II (2R) = 14.3567 μm; and height of the RVE II (H) = $2R\sqrt{3}$ μm.

### 4.1. Comparisons with numerical results

In order to verify the validity of the Mori-Tanaka model, the CNTs and the matrix material of the nanocomposite studied by Liu and Chen (2003) are considered for the constituent phases of the CNRP material. Table 6 illustrates this comparison and it may be observed that the two sets of results are in excellent agreement and thus validating the Mori-Tanaka model employed in this study. The effective elastic coefficients of the CNRP were computed by employing the Mori-Tanaka method. Obtained effective elastic constants of the CNRP are summarized in Table 7. It should be noted that the effective elastic coefficients of the CNRP were determined with respect to its RVE in which the maximum volume fraction of the CNT ($v_n$) was used. As expected, randomly dispersed CNTs provide the isotropic elastic properties for the resulting nanocomposite. Summarized results clearly demonstrate that the randomly dispersed CNTs improve the effective axial elastic coefficient ($C_{11}^c$) and shear modulus ($C_{44}^c$) of the nanocomposite over those of the transverse elastic coefficient ($C_{22}^c$) and the shear moduli ($C_{44}^c$ and $C_{66}^c$) of the nanocomposite reinforced with aligned CNTs. This is attributed to the fact that the CNTs are homogeneously dispersed in the epoxy matrix in the random case and hence the overall elastic properties of the resulting nanocomposite improve in comparison to the aligned case. These findings are consistent with the previously reported findings (Odegard et al., 2003; Wernik and Meguid, 2014).

Next, three-phase FE shear lag models were developed using the commercial software ANSYS 14.0 to confirm the applicability of the analytical shear lag model derived herein. With assumed hexagonal fiber packing array, a rectangular periodic RVE II was defined comprising the composite fibers and the polymer matrix. Note that only half of the RVE II was considered for the analysis taking advantage of symmetry. The geometry of the FE model, and the loading and boundary conditions were chosen such that they represent those of the actual experimental test. Also the geometric and material properties used in the FE simulations are identical to those of the analytical model. The microscale fiber, the CNRP and the epoxy matrix were constructed and meshed with twenty-node solid elements SOLID186. To analyze the stress transfer characteristics of the nano-reinforced composite, a tensile stress ($\sigma$) was applied to the RVE II along the x–direction at x = L and its one end was constrained at x = 0. Under the conditions of an imposed tensile stress ($\sigma$) on the RVE II of the nano-reinforced composite, the average stresses $\{\overline{\sigma}^k\}$ in the $k^{th}$ phase of the RVE II can be obtained as follows:

$$\{\overline{\sigma}^k\} = \frac{1}{V^k} \int \{\sigma^k\} \, dV^k \; ; \quad k = f, c \text{ and } m \tag{51}$$

where $V^k$ represents the volume of the $k^{th}$ phase of the RVE II and the field variable with an overbar represents the average of the field variable. For analyzing the stress transfer characteristics, the following non-dimensional parameters are used:

$$\sigma^* = \frac{\overline{\sigma}_x^f}{\sigma} \text{ and } \tau^* = \frac{\tau_1}{\sigma} \tag{52}$$

It should be noted that the FE shear lag model accounts for the interaction among the neighboring phases and provides the overall load transfer characteristics of the nano-reinforced composite. Tension-shear model with the regular staggering has been introduced by Gao and coworkers [55, 61] to interpret many underlying mechanisms in biocomposites. According to Gao and coworkers, the axial tensile stress ($\sigma$) imposed on the RVE is related to the shear stress ($\tau$) along the RVE I as shown in Fig 8 is given by

$$\sigma = \frac{L}{2R} v_{CF} \tau \tag{54}$$

in which $v_{CF}$ is volume fraction of the composite fiber in the RVE I. We consider the 2$^{nd}$ case (i.e., CNTs are aligned radially to the axis of the carbon fiber) to validate the analytical model developed herein. The comparisons of the average axial stress in the carbon fiber ($\sigma^*$) and the interfacial shear stress along its length ($\tau^*$) computed by the analytical and FE shear lag models are presented in Figs. 9 and 10, respectively. It should be noted that because of symmetry, distributions of the stresses in the zone of the carbon fiber reinforcement are plotted for one half of its length. It may be importantly observed from these figures that if the staggering effect of the adjacent RVEs I is neglected (that is, $\tau = 0$), then the analytical shear lag model over-predict the values of $\sigma^*$ and $\tau^*$ compared to those of the shear lag models predictions considering the staggering effect (that is, $\tau \neq 0$). Hence, the existence of the non-negligible shear tractions along the length of the RVE of the nano-reinforced composite cannot be ignored; they play a crucial role in the stress transfer characteristics of the nano-reinforced composite. It may also be observed that the good agreement between the two sets of results have been obtained and thus verifying the reliability of the present analytical shear lag model incorporating the staggering effect of the adjacent RVEs. We also validated the analytical results of the other two cases of CNT orientations against FE results and found them in good agreement.

## 4.2. Analytical shear lag model results

In this Section, parametric results of a shear lag analysis of the nano-reinforced composite have been presented to investigate the effects of CNT orientations, applied radial load (q), and the spacings between the adjacent composite fibers. First we present the effect of CNT orientations on the stress transfer characteristics of nano-reinforced composite. The considered three cases represent the practical situation of distribution of CNTs in the epoxy matrix. Figures 11 and 12 demonstrate that the CNT orientations significantly influence the values of $\sigma^*$ and $\tau^*$. It may be observed that the maximum values of $\sigma^*$ and $\tau^*$ are significantly reduced when the CNTs are randomly and axially aligned, respectively, compared to radially aligned case. Note that these results are presented without the application of radial load on the cylindrical RVE of nano-reinforced composite. The consideration of the radial load on the RVE accounts for the lateral extensional interaction between the adjacent composite fibers, which may arise due to the manufacturing processes. Therefore, the application of the radial load on the RVE was considered to focus on the trade-off between the randomly and axially aligned cases. The variations of the values of $\sigma^*$ and $\tau^*$ are presented in Figs. 13 and 14 when the value of q = $0.1\sigma$. It may be observed that the maximum values of $\sigma^*$ and $\tau^*$ are significantly decreased with the application of radial load. More importantly, the stress transfer characteristics of nano-reinforced composite are improved when the CNTs are randomly aligned. This is attributed to the fact that the stress transfer due to the 3D loading occurs from the matrix to the CNRP, the fiber, and under such situation, isotropic CNRP exhibits better stress transfer. Note that the overall elastic properties of CNRP improve when CNTs are homogeneously dispersed in it (see Table 7), which, in turn, improve the stress transfer characteristics of the resulting composite. The axially aligned case falls in between other two cases and enhances the stress transfer characteristics of the nano-reinforced composite over the radially aligned case. The subsequent results for investigating the effect of spacings between the adjacent composite fibers along their radial and axial directions on the stress transfer characteristics are presented considering the randomly dispersed case.

So far, in this work, the stress transfer characteristics of the nano-reinforced composite have been studied considering the values of the geometrical parameters R/b and $L/L_f$ as 1.1. Here, the geometrical parameters R/b and $L/L_f$ represent the spacings between the adjacent composite fibers along their radial and axial directions, respectively, over the volume of the composite lamina. Practically, the gaps between the adjacent composite fibers interlaced in the polymer matrix can vary over the volume of the composite lamina. The variation of such gaps for a particular value of $v_f$ would be an important study. For this the four discrete values of R/b and $L/L_f$ are considered as 1.025, 1.05, 1.075 and 1.1. Keeping the diameter of the carbon fiber constant as $2a = 10$ μm, the effective elastic coefficients of the CNRP containing randomly dispersed CNTs corresponding to these geometrical parameters are determined and listed in Table 8. The variations of the values of $\sigma^*$ and $\tau^*$ are presented in Figs. 15 and 16 for different values of $R/b$ and $L/L_f$ when the radial load is $q = 0.1\sigma$. The maximum value of $\sigma^*$ decreases with the increase in the values of $R/b$ and $L/L_f$. This is attributed to the fact that the effective elastic properties of the CNRP are improved and the staggering effect of the adjacent RVEs is increased with the increase in the spacings which eventually enhance the load carrying capacity of the composite. The maximum value of $\tau^*$ is found to be marginally affected by the variation of the values of $R/b$ and $L/L_f$. Although not shown here, the maximum values of $\sigma^*$ and $\tau^*$ are significantly decreased with the application of the radial load on the RVE.

## 5. Conclusions

In this work, we developed a multiscale model to investigate the stress transfer characteristics of a nano-reinforced composite considering the staggering effect. The distinctive feature of the construction of this composite is that the introduction of CNTs around the microscale carbon fibers embedded in the epoxy matrix, resulting in a nano-reinforced composite with enhanced properties. Using this concept, damage initiation threshold and the fatigue strength of conventional composites can be greatly improved by toughening the interfacial fiber-matrix region. Accordingly, two aspects were examined. First, orthotropic effective elastic properties of CNT-reinforced polymer nanocomposite (CNRP) were determined through MD simulations in conjunction with the Mori-Tanaka model. Second, three-phase analytical and numerical models for the nano-reinforced composite were developed. Predictions of an analytical model have also been validated with those of FE simulations. The developed analytical model was then applied to investigate the effect of orientations of CNTs on the stress transfer characteristics of the nano-reinforced composite. Orientation of CNTs has significant influence on the stress transfer characteristics of the composite; randomly dispersed CNTs around the microscale fiber was found be effective in comparison to all other orientations. Moreover, the stress transfer characteristics of the nano-reinforced composite are significantly improved with the increase in the spacings between the carbon fibers irrespective of the magnitude of the radial load. The multiscale model developed herein offers significant advantages over the existing shear lag models and is capable of investigating the stress transfer characteristics of any multiscale composite considering staggering effect and three dimensional loads.

**Acknowledgements:** This work was supported by a Banting Postdoctoral Fellowship awarded to the first author from the Natural Sciences and Engineering Research Council of Canada (NSERC).

# Appendix A

The elements of the Eshelby tensor $[S^E]$ [62]:

$$S_{1111}^E = S_{2222}^E = \frac{5 - 4\nu^m}{8(1 - \nu^m)}, \qquad S_{1122}^E = S_{2211}^E = \frac{4\nu^m - 1}{8(1 - \nu^i)},$$

$$S_{1133}^E = S_{2233}^E = \frac{\nu^m}{2(1 - \nu^m)}, \quad S_{1313}^E = S_{2323}^E = 1/4,$$

$$S_{1212}^E = \frac{3 - 4\nu^m}{8(1 - \nu^m)} \text{ and all other elements are zero} \tag{A.1}$$

Direction cosines corresponding to the transformation matrix [Eq. (5)]:

$$t_{11} = \cos\phi \, \cos\psi - \sin\phi \, \cos\gamma \, \sin\psi, \ t_{12} = \sin\phi \, \cos\psi + \cos\phi \, \cos\gamma \, \sin\psi,$$

$$t_{13} = \sin\psi \, \sin\gamma \, , \ t_{21} = -\cos\phi \, \sin\psi - \sin\phi \, \cos\gamma \, \cos\psi,$$

$$t_{22} = -\sin\phi \, \sin\psi + \cos\phi \, \cos\gamma \, \cos\psi, \ t_{23} = \sin\gamma \, \cos\psi, \ t_{31} = \sin\phi \, \sin\gamma,$$

$$t_{32} = -\cos\phi \, \sin\gamma \quad \text{and} \quad t_{33} = \cos\gamma \tag{A.2}$$

# Appendix B

The constants $(A_i)$ obtained in the course of deriving the shear lag model in Section 4 are explicitly expressed as follows:

$$A_1 = \frac{a}{C_{66}^c(b^2 - a^2)}\left[b^2\ln\frac{r}{a} - \frac{(r^2 - a^2)}{2}\right], \qquad A_2 = -\frac{b}{C_{66}^c(b^2 - a^2)}\left[a^2\ln\frac{r}{a} - \frac{(r^2 - a^2)}{2}\right]$$

$$A_3 = \frac{a}{C_{66}^c(b^2 - a^2)}\left[b^2\ln\frac{b}{a} - \frac{(b^2 - a^2)}{2}\right]$$

$$A_4 = \frac{b}{C_{66}^m(R^2 - b^2)}\left[R^2\ln\frac{r}{b} - \frac{(r^2 - b^2)}{2}\right] - \frac{b}{C_{66}^c(b^2 - a^2)}\left[a^2\ln\frac{b}{a} - \frac{(b^2 - a^2)}{2}\right]$$

$$A_5 = b^2\ln\frac{b}{a} - \frac{(b^2 - a^2)}{2}, \qquad A_6 = a^2\ln\frac{b}{a} - \frac{(b^2 - a^2)}{2},$$

$$A_7 = \frac{b}{C_{66}^c(b^2 - a^2)}\left[a^2\ln\frac{b}{a} - \frac{(b^2 - a^2)}{2}\right]$$

$$A_8 = \frac{b}{C_{66}^m(R^2 - b^2)}\left[R^2\ln\frac{R}{b} - \frac{(R^2 - b^2)}{2}\right] - \frac{b}{C_{66}^c(b^2 - a^2)}\left[a^2\ln\frac{b}{a} - \frac{(b^2 - a^2)}{2}\right]$$

$$A_9 = -\left[\frac{2a^2 C_{12}^c}{b^2 - a^2} + \frac{2C_{11}^c C_{12}^f}{C_{11}^f}\right], \qquad A_{10} = \frac{2b^2 C_{12}^c}{b^2 - a^2}, \qquad A_{11} = \frac{2C_{12}^c}{b^2 - a^2}$$

$$A_{12} = \frac{aC_{11}^c}{C_{66}^c(b^2 - a^2)^2}\left[a^2 b^2 + b^4\ln\frac{b}{a} - \frac{a^4}{4} - \frac{3b^4}{4}\right],$$

$$A_{13} = -\frac{bC_{11}^c}{C_{66}^c(b^2 - a^2)^2}\left[a^2 b^2\ln\frac{b}{a} + \frac{a^4}{4} - \frac{b^4}{4}\right], \qquad A_{14} = \frac{C_{11}^c}{C_{11}^f} + A_9 b_{11} + A_{10} b_{21} + A_{11} b_{31},$$

$$A_{15} = A_9 b_{12} + A_{10} b_{22} + A_{11} b_{32} + A_{12}, \qquad A_{16} = A_9 b_{13} + A_{10} b_{23} + A_{11} b_{33} + A_{13},$$

$$A_{17} = A_9 b_{14} + A_{10} b_{24} + A_{11} b_{34}, \qquad A_{18} = A_9 b_{15} + A_{10} b_{25} + A_{11} b_{35},$$

$$A_{19} = \frac{1}{C_{66}^m(R^2 - b^2)^2}\left[b^2 R^2 + R^4\ln\frac{R}{b} - \frac{b^4}{4} - \frac{3R^4}{4}\right] - \frac{1}{C_{66}^c(b^2 - a^2)}\left[a^2\ln\frac{b}{a} - \frac{(b^2 - a^2)}{2}\right],$$

$$A_{20} = \frac{C_{11}^m}{C_{11}^f} - \frac{2C_{11}^m C_{12}^f}{C_{11}^f}b_{11} + 2C_{12}^m b_{21}, \qquad A_{21} = -\frac{2C_{11}^m C_{12}^f}{C_{11}^f}b_{12} + 2C_{12}^m b_{22} + C_{11}^m A_3,$$

$$A_{22} = -\frac{2C_{11}^m C_{12}^f}{C_{11}^f}b_{13} + 2C_{12}^m b_{23} + C_{11}^m b A_{19}, A_{23} = -\frac{2C_{11}^m C_{12}^f}{C_{11}^f}b_{14} + 2C_{12}^m b_{24},$$

$$A_{24} = -\frac{2C_{11}^m C_{12}^f}{C_{11}^f} b_{15} + \frac{RC_{11}^m}{C_{66}^m(R^2 - b^2)}\left[R^2 \ln\frac{R}{b} - \frac{(R^2 - b^2)}{2}\right] + 2C_{12}^m b_{25},$$

$$A_{25} = a^2 + (b^2 - a^2)A_{14} + (R^2 - b^2)A_{20}, \qquad A_{26} = -(a/2)[(b^2 - a^2)A_{15} + (R^2 - b^2)A_{21}],$$

$$A_{27} = \left(\frac{R^2 - b^2}{2b}\right)[(b^2 - a^2)A_{16} + (R^2 - b^2)A_{22}], \qquad A_{28} = (b^2 - a^2)A_{17} + (R^2 - b^2)A_{23},$$

$$A_{29} = (b^2 - a^2)A_{18} + (R^2 - b^2)A_{24}, \qquad A_{30} = -\frac{a}{2}\left[\frac{A_{15}A_{22}}{A_{16}} - A_{21}\right],$$

$$A_{31} = \frac{1}{A_{30}}\left[\frac{A_{14}A_{22}}{A_{16}} - A_{20} + \left(\frac{a^2}{b^2 - a^2}\right)\frac{A_{22}}{A_{16}} - \left(\frac{R^2 - b^2}{b^2 - a^2}\right)\frac{A_{22}A_{25}}{A_{16}A_{27}} - \frac{A_{26}}{A_{27}}\right],$$

$$A_{32} = -\frac{1}{A_{30}}\left[\left(\frac{R^2 - b^2}{b^2 - a^2}\right)\frac{A_{22}A_{25}}{A_{16}A_{26}} + \frac{A_{25}}{A_{27}}\right], \qquad A_{33} = -\frac{1}{A_{30}}\left[\left(\frac{R^2 - b^2}{b^2 - a^2}\right)\frac{R^2 A_{22}}{A_{16}A_{27}} + \frac{R^2}{A_{27}}\right],$$

$$A_{34} = -\frac{1}{A_{30}}\left[\left(\frac{R^2 - b^2}{b^2 - a^2}\right)\frac{A_{22}A_{28}}{A_{16}A_{27}} + \frac{A_{28}}{A_{27}}\right] \text{ and } A_{35} = -\frac{1}{A_{30}}\left[\left(\frac{R^2 - b^2}{b^2 - a^2}\right)\frac{A_{22}A_{29}}{A_{16}A_{27}} + \frac{A_{29}}{A_{27}}\right]$$

**Table 1** Parameters used in the MD unit cells

| Parameter | Pure epoxy | Unit cell to determine RDF | Nano-fiber |
|---|---|---|---|
| CNT type | – | (9, 9) | (9, 9) |
| CNT diameter (Å) | – | 12.2 | 12.2 |
| CNT length (Å) | – | 43 | 43 |
| Unit cell dimensions (Å$^3$) | 39×39×39 | 50×50×43 | 31×31×43 |
| CNT volume fraction | – | 5.5% | 12.16% |
| Total number of atoms | 6250 | 12690 | 4340 |

**Table 2** Elastic moduli of the epoxy material

| | Young's modulus (GPa) | Poisson's ratio |
|---|---|---|
| Present MD simulations | 3.5 | 0.36 |
| Experimental [47] | 3.43 | - |

**Table 3** Effective elastic coefficients of the RVEs and corresponding displacement fields

| Elastic coefficients | Applied strains | Applied displacement |
|---|---|---|
| $C_{11}$ | $\varepsilon_{11} = e$ | $u_1 = ex_1$ |
| $C_{33}$ | $\varepsilon_{33} = e$ | $u_3 = ex_3$ |
| $C_{44}$ | $\varepsilon_{23} = e/2$ | $u_2 = \frac{e}{2}x_3,$ $u_3 = \frac{e}{2}x_2$ |
| $C_{66}$ | $\varepsilon_{12} = e/2$ | $u_1 = \frac{e}{2}x_2,$ $u_2 = \frac{e}{2}x_1$ |
| $K_{12} = \dfrac{C_{11} + C_{12}}{2}$ | $\varepsilon_{11} = \varepsilon_{22}$ | $u_1 = ex_1,$ $u_2 = ex_2$ |

**Table 4** Material properties of the nano-fiber

| CNT volume fraction | $C_{11}$ (GPa) | $C_{12}$ (GPa) | $C_{23}$ (GPa) | $C_{33}$ (GPa) | $C_{44}$ (GPa) | $C_{66}$ (GPa) |
|---|---|---|---|---|---|---|
| 12.16% | 15.1 | 7.5 | 6 | 116.4 | 2 | 3.8 |

**Table 5** Material properties of the constituent phases of the composite

| Material | $C_{11}$ (GPa) | $C_{12}$ (GPa) | $C_{13}$ (GPa) | $C_{23}$ (GPa) | $C_{33}$ (GPa) | $C_{44}$ (GPa) | $C_{66}$ (GPa) | (nm) |
|---|---|---|---|---|---|---|---|---|
| Carbon fiber (Honjo, 2007) | 236.4 | 10.6 | 10.6 | 10.7 | 24.8 | 7 | 25 | d = 10000 |
| Polyimide (Odegard et al. 2005) | 9 | 6 | 6 | 6 | 9 | 1.5 | 1.5 | - |

**Table 6** Comparison of the elastic constants of the CNRP material

| $E^n/E^p$ | $E_1/E^p$ | | $E_2/E^p$ | | $\nu_{12}, \nu_{13}$ | |
|---|---|---|---|---|---|---|
| | FEM | Mori-Tanaka | FEM | Mori-Tanaka | FEM | Mori-Tanaka |
| 5 | 1.1948 | 1.1948 | 1.1737 | 1.0666 | 0.3 | 0.3 |
| 10 | 1.4384 | 1.4384 | 1.3336 | 1.0912 | 0.3 | 0.3 |

$E^n = 1000$ GPa, $\nu^n = 0.3, \nu^p = 0.3$ and CNT volume fraction, $v_n = 0.04871$ Liu and Chen (2003); where $E^n$ and $E^p$ are the Young's moduli of the CNT and the polymer matrix, respectively; $E_1$ and $E_2$ are the axial and the transverse Young's moduli of the unwound PMNC, respectively; $\nu^n$ and $\nu^p$ are the Poisson's ratios of the CNT and the polymer matrix, respectively; $\nu_{12}$ and $\nu_{23}$ are the axial and the transverse Poisson's ratios of the unwound PMNC, respectively.

**Table 7** Effective elastic coefficients of the CNRP

| Orientation of CNTs | $V_{CNT}$ | $v_n$ | $C_{11}^c$ (GPa) | $C_{12}^c$ (GPa) | $C_{23}^c$ (GPa) | $C_{33}^c$ (GPa) | $C_{44}^c$ (GPa) | $C_{66}^c$ (GPa) |
|---|---|---|---|---|---|---|---|---|
| Axially Aligned | 0.0297 | 0.1055 | 102.18 | 6 | 7.29 | 13.96 | 6.67 | 2.66 |
| Radially Aligned | 0.0297 | 0.1055 | 13.90 | 6.67 | 18.05 | 46 | 27.95 | 1.93 |
| Randomly Dispersed | 0.0297 | 0.1055 | 29 | 13.05 | 13.05 | 29 | 7.9 | 7.9 |

**Table 8** Effective elastic coefficients of the CNRP containing randomly dispersed CNTs

| $R/b = L/L_f$ | $V_{CNT}$ | $v_n$ | $C_{11}^c$ (GPa) | $C_{12}^c$ (GPa) | $C_{23}^c$ (GPa) | $C_{33}^c$ (GPa) | $C_{44}^c$ (GPa) | $C_{66}^c$ (GPa) |
|---|---|---|---|---|---|---|---|---|
| 1.025 | 0.0439 | 0.0992 | 28 | 12.6 | 12.6 | 28 | 7.7 | 7.7 |
| 1.05 | 0.0389 | 0.1014 | 28.3 | 12.7 | 12.7 | 28.3 | 7.8 | 7.8 |
| 1.075 | 0.0341 | 0.1034 | 28.6 | 12.9 | 12.9 | 28.6 | 7.85 | 7.85 |

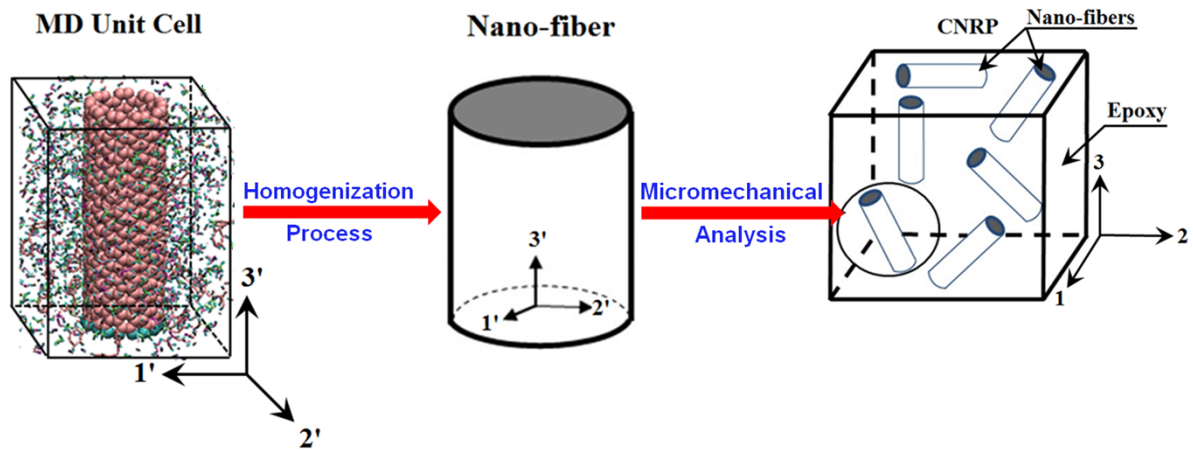

**Fig. 1.** Modeling steps involved in the developed multiscale model.

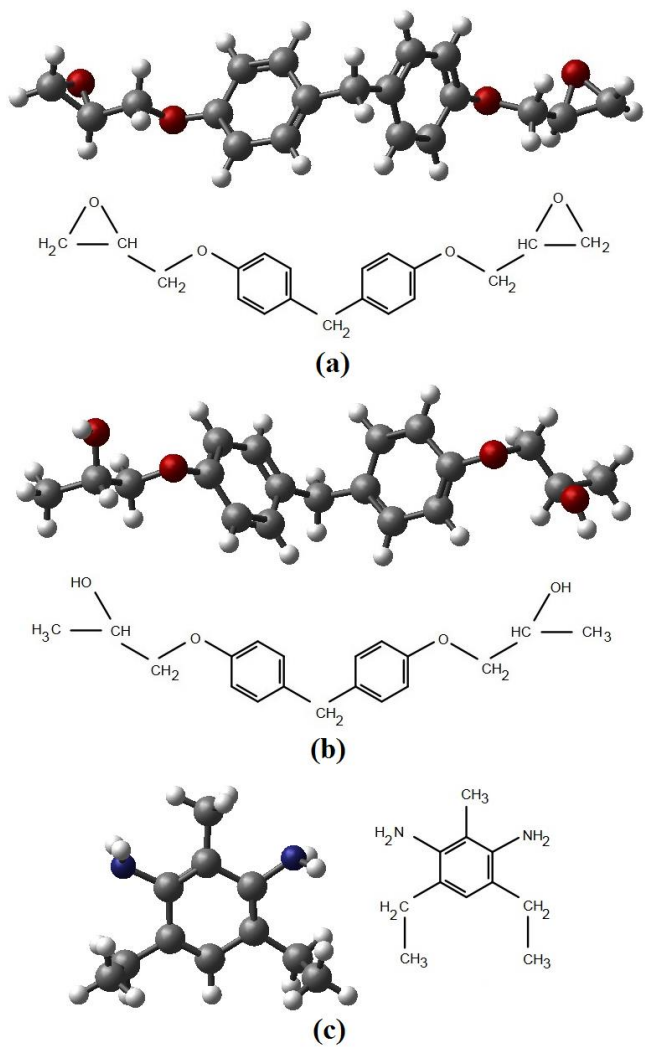

**Fig. 2.** Molecular and chemical structures of (a) EPON 862, (b) activated EPON 862 and (c) DETDA curing agent. Colored with gray, red, blue, and white are carbon, oxygen, nitrogen, and carbon atoms, respectively.

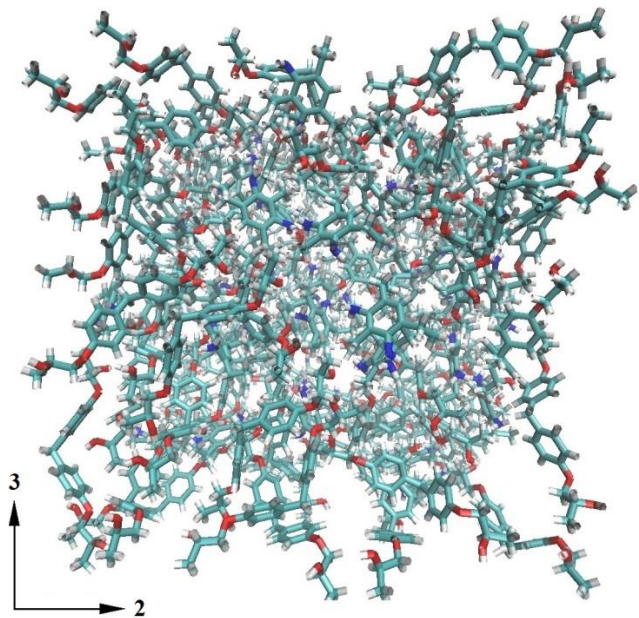

**Fig. 3.** Snapshot of atomic configurations of EPON 862-DETDA epoxy.

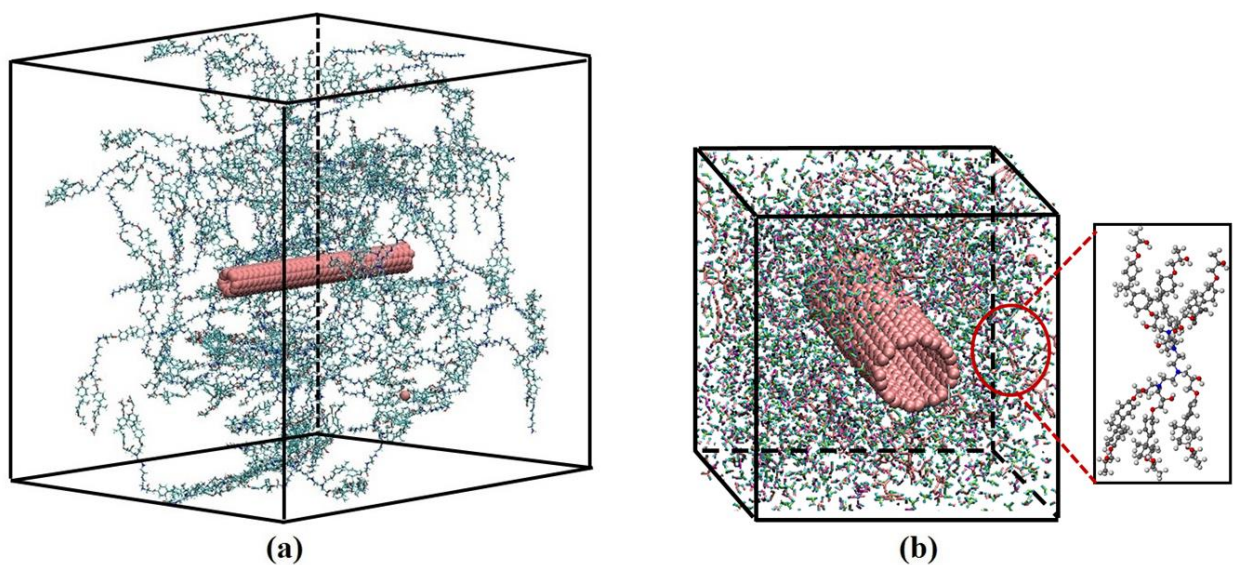

**(a)**                                    **(b)**

**Fig. 4.** The simulation box used to study the interface layer: (a) before volume reduction, and (b) after equilibration.

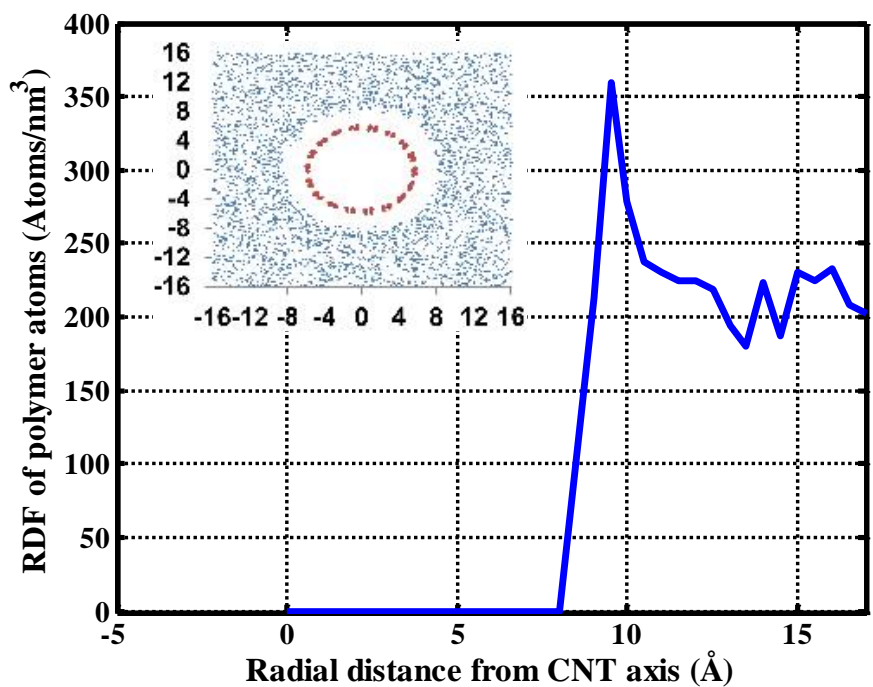

**Fig. 5.** RDF of the epoxy atoms around the embedded CNT (transverse cross-sectional view of the equilibrated unit cell is shown in the inset).

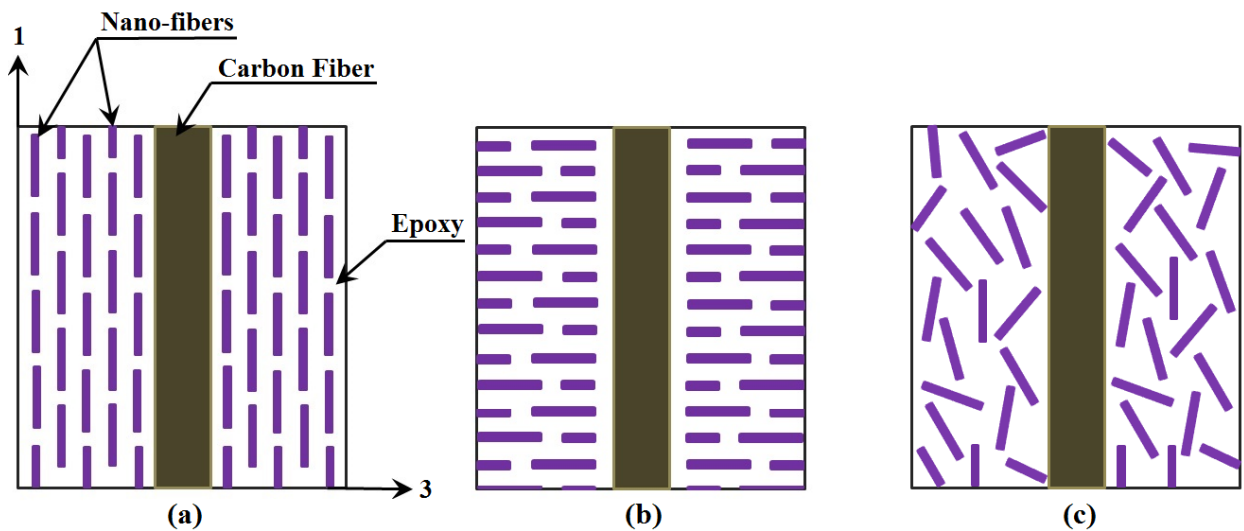

**Fig. 6.** Schematic cross-sections of the composite fibers in which nano-fibers are: (i) aligned along the direction of the fiber, (ii) aligned radially to the axis of the fiber, and (iii) randomly dispersed.

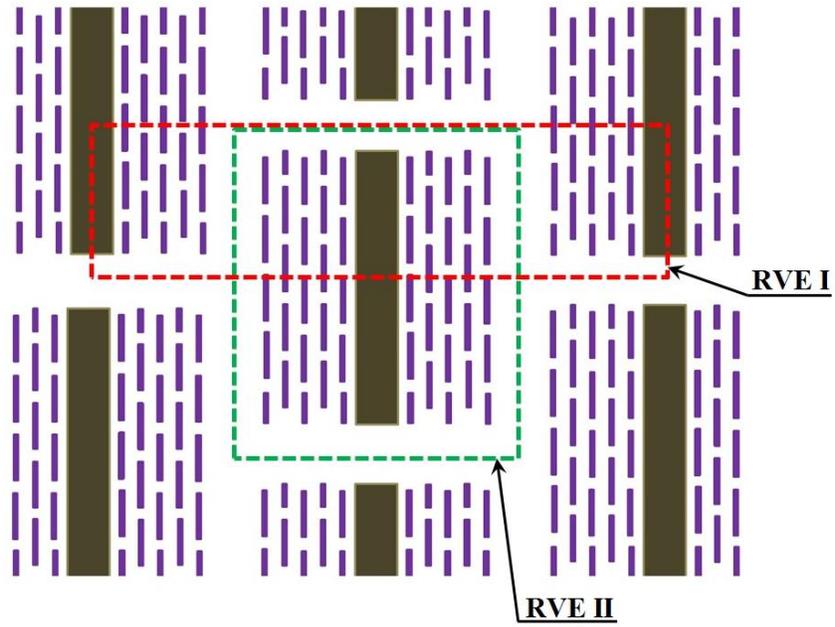

**Fig. 7.** In-plane cross-section of the nano-reinforced composite lamina.

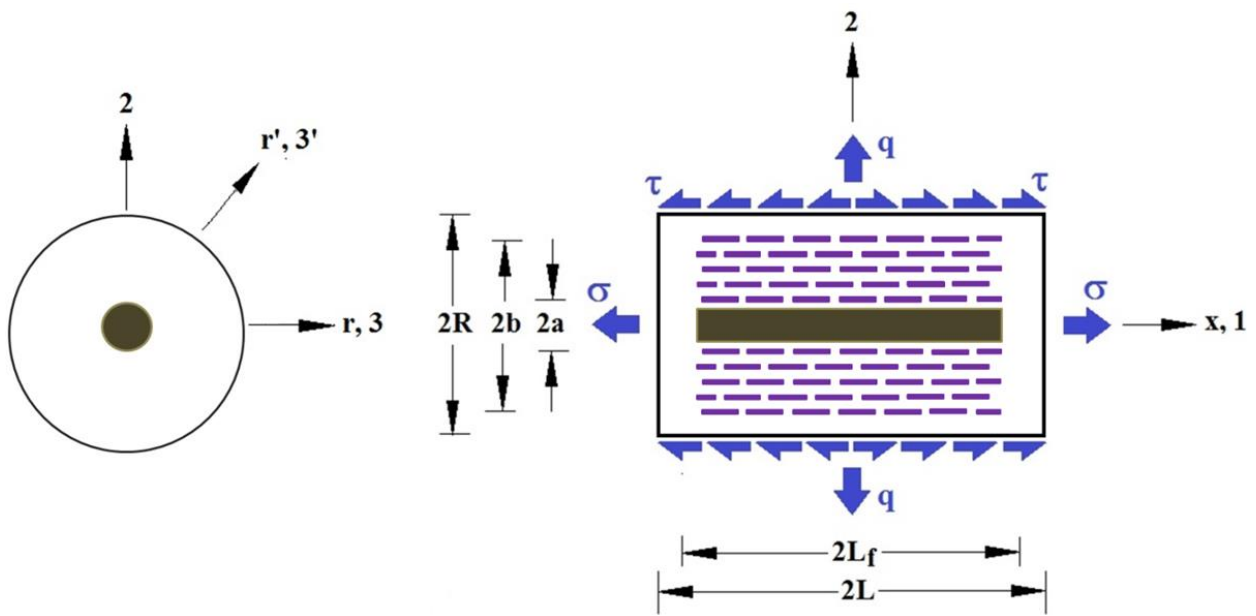

**Fig. 8.** Cylindrical RVE I of the nano-reinforced composite subjected to axial ($\sigma$) and radial ($q$) loads.

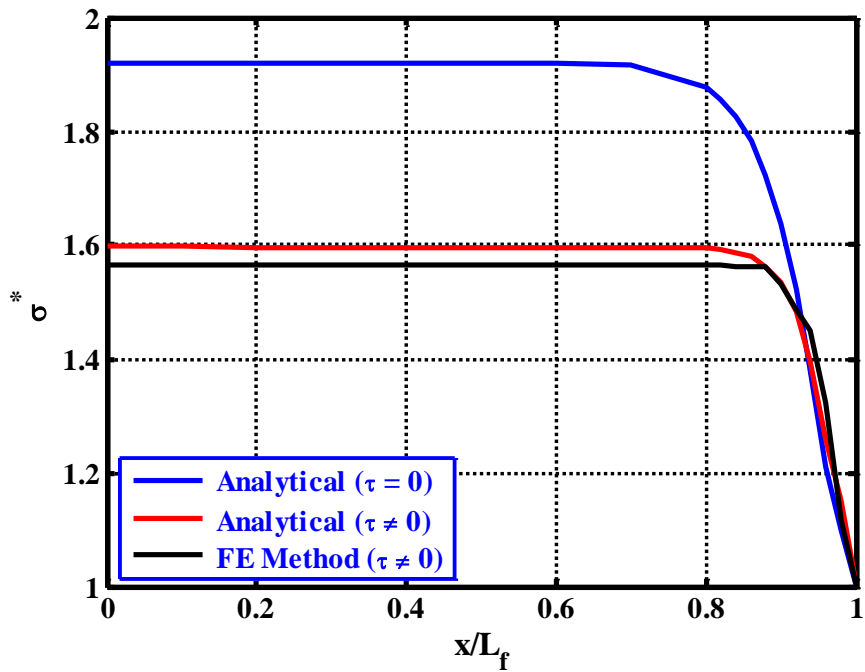

**Fig. 9.** Comparison of the axial stress in the carbon fiber along its length predicted by the analytical and by FE shear lag models (R/b = L/L_f = 1.1, q = 0).

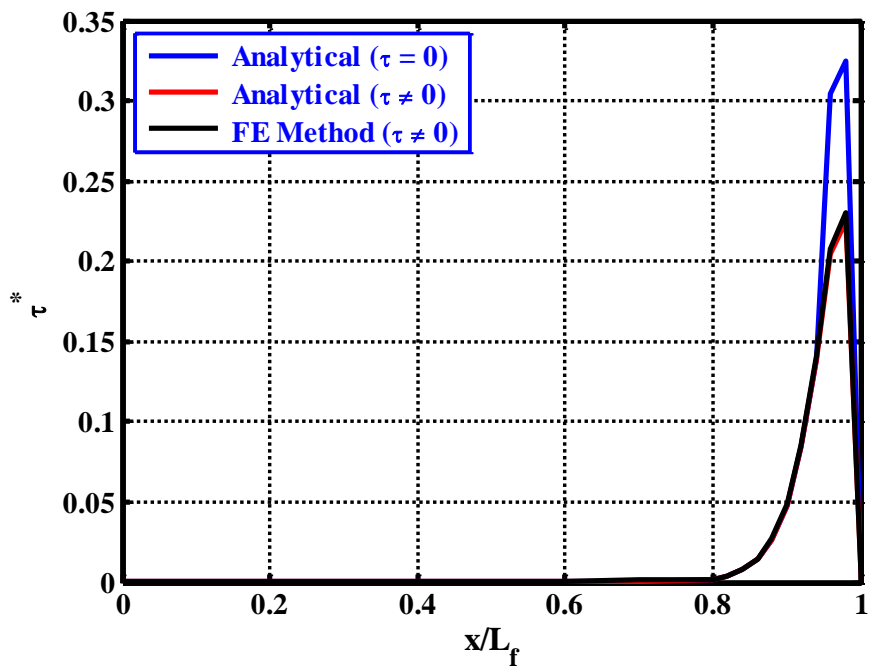

**Fig. 10.** Comparison of the interfacial shear stress along the length of carbon fiber predicted by the analytical and FE shear lag models (R/b = L/L_f = 1.1, q = 0).

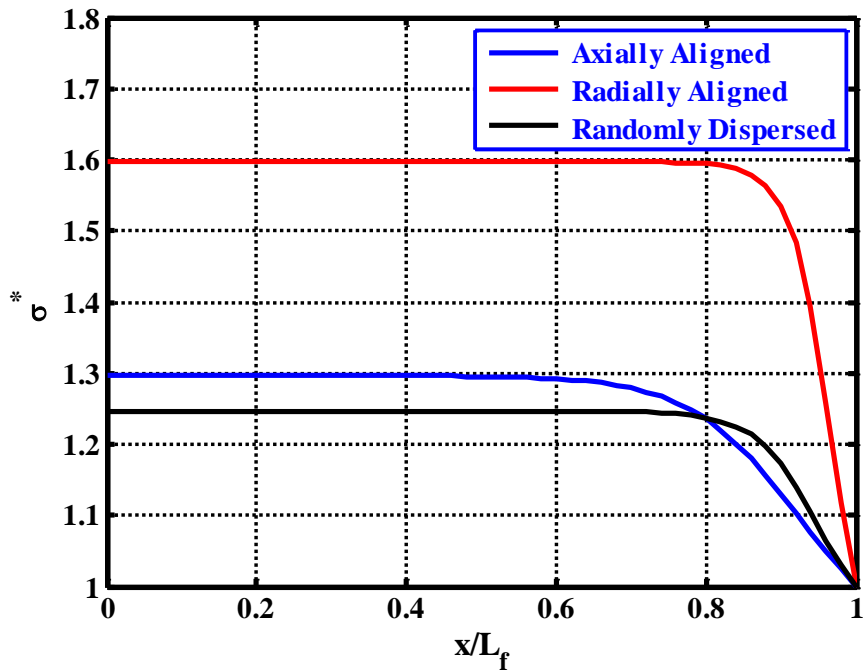

**Fig. 11.** Variation of the axial stress in the carbon fiber along its length for different orientations of CNTs ($q = 0$, $R/b = L/L_f = 1.1$).

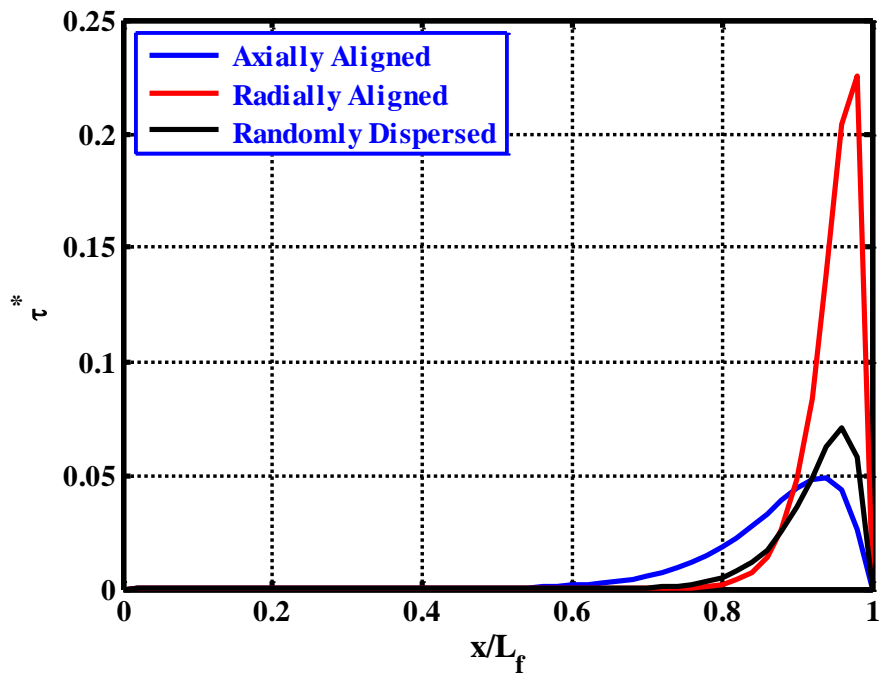

**Fig. 12.** Variation of the interfacial shear stress along the length of carbon fiber for different different orientations of CNTs ($q = 0$, $R/b = L/L_f = 1.1$).

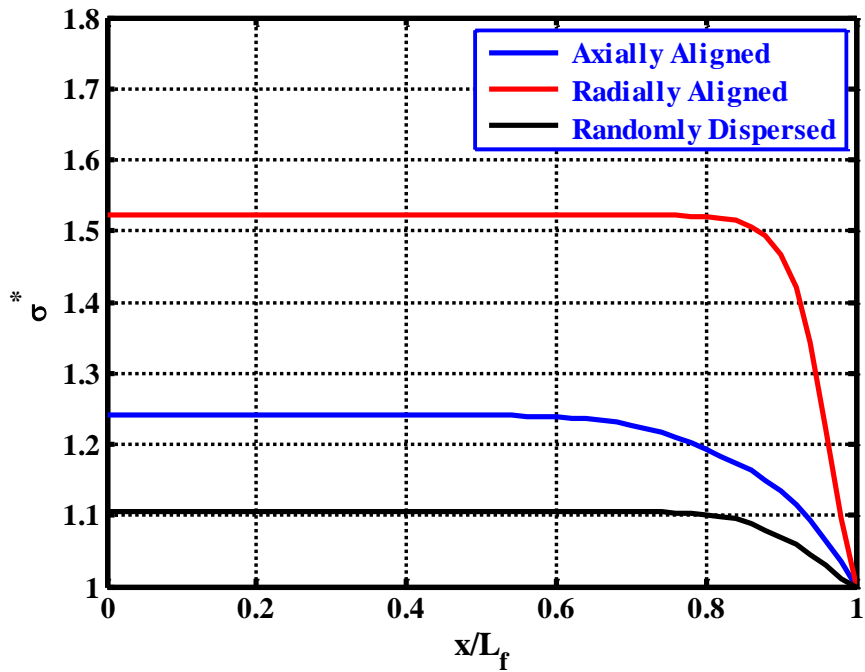

**Fig. 13.** Variation of the axial stress in the carbon fiber along its length for different orientations of CNTs
$(q/\sigma = 0.1, R/b = L/L_f = 1.1)$.

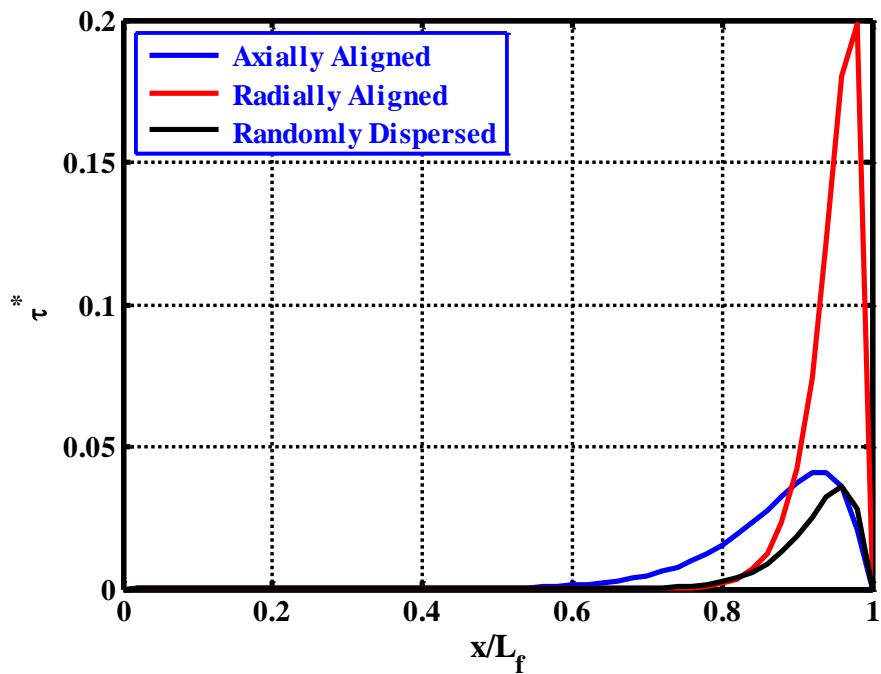

**Fig. 14.** Variation of the interfacial shear stress along the length of carbon fiber for different orientations
of CNTs $(q/\sigma = 0.1, R/b = L/L_f = 1.1)$.

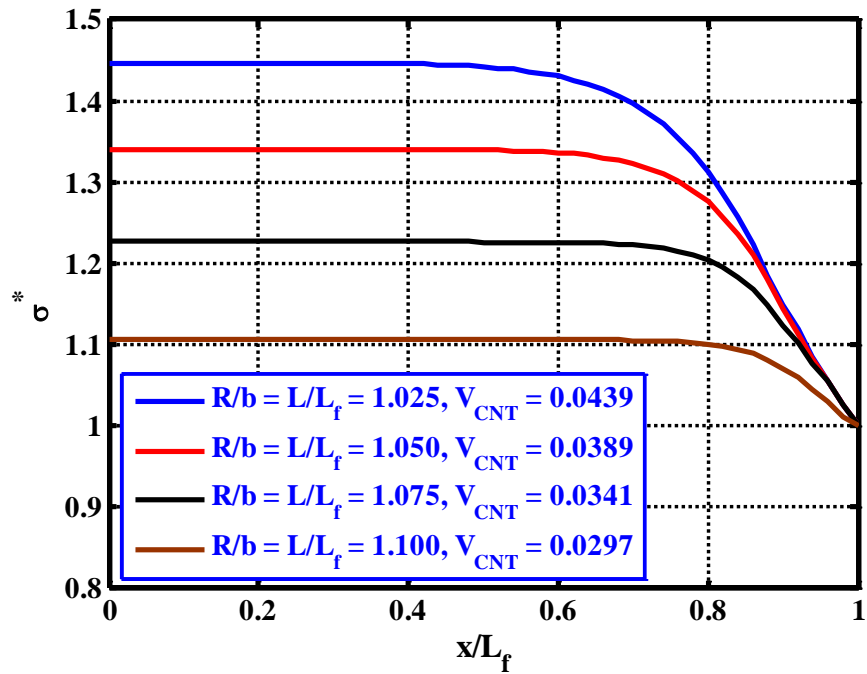

**Fig. 15.** Variation of the axial stress in the carbon fiber along its length for different values of R/b and $L/L_f$ (q = 0.1).

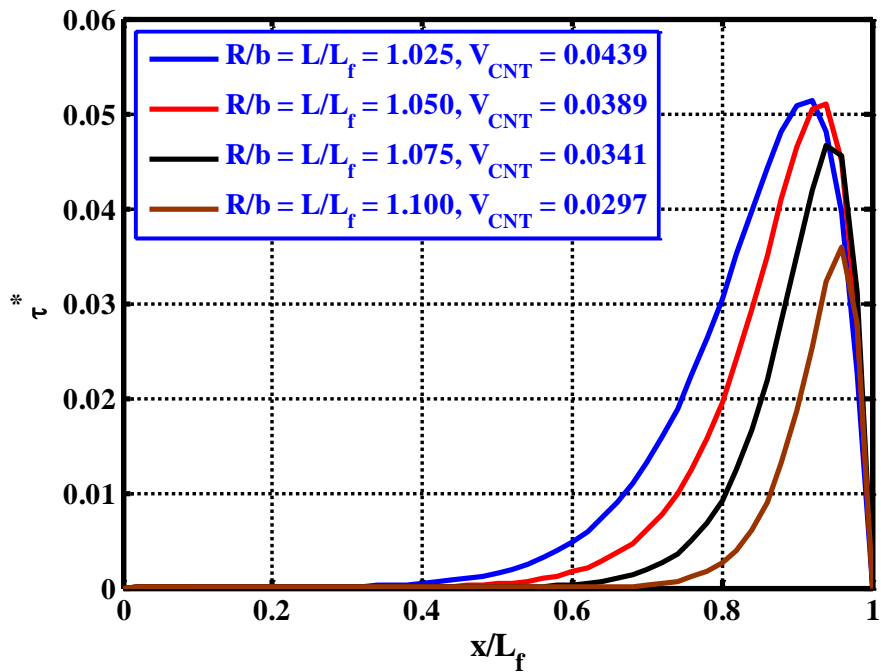

**Fig. 16.** Variation of the interfacial shear stress along the length of carbon fiber for different values of R/b and $L/L_f$ (q = 0.1).